\documentclass[prb,twocolumn,lscape]{revtex4}
\usepackage{amsmath,graphicx,epsfig}
\usepackage{colordvi}
\newcommand{\kb}{k_{B}}
\newcommand{\imag}{\mbox{Im}}
\newcommand{\real}{\mbox{Re}}
\newcommand{\trace}{\mbox{Tr}}
\newcommand{\lrpart}[1]{\stackrel{\leftrightarrow}{\partial}_{#1}}
\newcommand{\rpart}[1]{\overrightarrow{\partial_{#1}}}
\newcommand{\lpart}[1]{\overleftarrow{\partial_{#1}}}
\newcommand{\grad}[1]{\vec{\nabla}\negthinspace_{#1}}
\newcommand{\cellvol}{\mbox{V}\negthinspace_{\mbox{c}}}

\newcommand{\ra}{\rangle }
\newcommand{\la}{\langle }

\newcommand{\iomn}{i\omega_n}

\begin{document}

\title{Towards realistic electronic structure
calculations of strongly correlated electron systems}

\author{V. S. Oudovenko}
\affiliation{Bogoliubov Laboratory for Theoretical Physics, %
Joint Institute for Nuclear Research, 141980 Dubna, Russia}
\affiliation{Center for Materials Theory, Department of Physics
and Astronomy, Rutgers University, Piscataway, NJ 08854}
\author{G. P\'alsson}
\affiliation{Center for Materials Theory, Department of Physics
and Astronomy, Rutgers University, Piscataway, NJ 08854}
\author{K. Haule}
\affiliation{Center for Materials Theory, Department of Physics
and Astronomy, Rutgers University, Piscataway, NJ 08854}
\author{S. Y. Savrasov}
\affiliation{Department of Physics, New Jersey Institute of
Technology, Newark, NJ 07102}
\author{G. Kotliar}
\affiliation{Center for Materials Theory, Department of Physics
and Astronomy, Rutgers University, Piscataway, NJ 08854}

\date{\today}

\begin{abstract}
We review some aspects of the realistic implementation of the
dynamical mean-field method. We extend the techniques introduced
in Ref.~\onlinecite{Georges:1996} to include the calculations of
transport coefficients. The approach is illustrated on
La$_{1-x}$Sr$_x$TiO$_3$ material undergoing a density driven Mott
transition.
\end{abstract}


\maketitle


\section{Introduction}%
\label{sec:introduction}

In recent years  understanding of the physics of strongly
correlated materials has undergone tremendous increase. This is
in part due to the advances in the theoretical treatments of
correlations, such as the development of dynamical mean-field
theory (DMFT) \cite{Georges:1996}.  The great allure of DMFT is
the flexibility of the method and its adaptability to different
systems as well as the simple conceptual picture it allows us to
form of the dynamics of the system.  The mean-field nature of the
method and the fact that the solution maps onto an impurity
model, many of which have been thoroughly studied in the past,
means that a great body of previous work can be brought to bear
on the solution of models of correlated lattice electrons. This
is exemplified by the great many numerical methods that can be
employed to solve the DMFT equations.

DMFT has been very successful in understanding the mechanism of
the Mott transition in model Hamiltonians. We now understand that
the various concentration induced phase transitions can be viewed
as bifurcation of a single functional of the Weiss field. The
phase diagram of the one-band Hubbard model, demonstrating that
there is a first order Mott transition at finite temperatures is
fully established ~\cite{Georges:1996}. Furthermore Landau like
analysis demonstrates that all the qualitative features are quite
generic at high temperatures ~\cite{Kotliar:1999}. However the
low-temperature ordered phases, and the quantitative aspects of
the spectra of specific materials clearly require realistic
treatment.

This triggered realistic development of DMFT in the last decade
which has now reached the stage that we can start tackling real
materials from an almost ab initio approach
\cite{Anisimov:1997,Lichtenstein:1998}, something which in the
past have been exclusively in the domain of density functional
theories.  We are now starting to see the merger of DMFT and such
ab initio techniques and consequently the opportunities for doing
real electronic structure calculations for strongly correlated
materials which so far were not within the reach of traditional
density functional theories.

Density functional theory (DFT)~\cite{DFTbook:1983} is canonical
example of ab initio approach, very successful in predicting
ground state properties of many systems which are less
correlated, for example the elemental metals and semiconductors.
However it fails in more correlated materials. It is unable to
predict that any systems is a Mott insulator in the absence of
magnetic order. It  is also not able to describe correctly
strongly correlated metallic state. As a matter of principle DFT
is a theory of ground state or thermodynamic properties at finite
temperatures. It's Kohn-Sham spectra cannot be rigorously
identified with the excitation spectra of the system. In weakly
correlated substances the Kohn-Sham spectra is a good
approximation to start a perturbative treatment of the
one-electron spectra using the GW method
\cite{Aryasetiawan:1998}. However this approach breaks down in
strongly correlated situations, because it is unable to produce
Hubbard bands. In orbitally ordered situations the LDA+U method
\cite{Anisimov:1991} produces the Hubbard bands, however this
method fails to produce quasiparticle bands and hence it is
unable to describe strongly correlated metals. Furthermore, once
long range order is lost the LDA+U method reduces to LDA and
hence it becomes inappropriate even for Mott insulators.

Dynamical mean-field theory is the simplest theory that is able
to describe on the same footing total energies and the spectra of
correlated electrons even when it contains both quasiparticle and
Hubbard bands. Combined with LDA, one then has a theory which
reduces to a successful method (LDA) in the weak correlations
limit. In the static limit, one can show ~\cite{Yang:2001} that
LDA+U can be viewed as a static limit of LDA+DMFT used in
conjunction with the Hartree-Fock approximation. Therefore LDA+U
is equivalent to LDA+DMFT+further approximations which are the
only justified in static ordered situations. Up to now, the
realistic LDA band structure was considered with DMFT for purpose
of computing one-electron (photoemission) spectra and total
energies.

Following Refs.~\onlinecite{Georges:1996, Anisimov:1997,
Lichtenstein:1998} in this paper we extend this approach to
computation of transport properties. Many transport studies
within DMFT applied to model Hamiltonians have been carried out,
and the strengths (non-perturbative character) and limitations
(absence of vertex corrections) are well understood. However
applications to real materials require realistic computations of
current matrix elements.

There are two ways in which DMFT can be used to understand the
physics of real materials. The simplest approach, outlined in
Ref.~\onlinecite{Georges:1996,Anisimov:1997} is closely tied to
the idea of model Hamiltonians. This requires i) methodology for
deriving of the hopping parameters and the interaction constants
ii) a technique for solving the dynamical mean-field equations
and iii) an algorithm for evaluating the transport function which
enters in the equations of transport coefficients. The second
direction is more ambitious and focus on an integration of i) and
ii) using functional formulations~\cite{Kotliar:2001}.

In this paper we review the first approach. The emphasis here is
in illustration of different aspects of the modeling which affect
the final answer. This is necessary to obtain a balanced approach
towards materials calculations. There are now many impurity
solvers, they differ in their accuracy and computational cost. In
the present paper we use two impurity solvers the Hirsch-Fye
Quantum Monte Carlo (QMC) method ~\cite{Hirsch:1986} and
symmetrized finite-U NCA method (SUNCA) \cite{Haule:2001}
comparing them in the context of simplified models without the
additional complications of real materials. Instead we use the
SUNCA  method as an impurity solver to compute the transport
properties and new developments using La$_{1-x}$Sr$_x$TiO$_3$ as
an example material ~\cite{Nekrasov:2000}. For other reviews of
realistic implementations of DMFT and electronic structure see
Ref.~\onlinecite{Held:2001}.

In the next section~\ref{sec:dmft} we shortly review a basic
dynamical mean-field theory concepts and their application to
realistic structure calculations.  As computation of transport
parameters requires knowledge of the self-energy coming from DMFT
calculations which is based on impurity solvers  we present a
short review of two impurity solvers used in the paper in
section~\ref{sec:imsolvers}.  Theory of the transport calculations
is given in section~\ref{sec:transport_theory}. Test system used
for transport calculations, which is doped LaTiO$_3$ ceramics,
and DMFT results are described in section~\ref{sec:dmft_results}.
Results of $dc$- transport calculations are presented in
section~\ref{sec:transp_results}. And finally we come to
conclusion in homonymous section~\ref{sec:conclusion}.

\section{Dynamical Mean-Field Theory}%
\label{sec:dmft}

\subsection{Realistic DMFT formalism}%

A central concept in electronic structure theory is the $f$-model
Hamiltonian. Conceptually, one starts from the full many body
problem containing all electrons and then proceeds to eliminate
some high-energy degrees of freedom. The results is a Hamiltonian
containing only a few bands. The determination of the model
Hamiltonian is a difficult problem in itself, which has received
a significant attention~\cite{Marel:1988, McMahan:1988,
Hybertsen:1989, Annett:1989, Gunnarsson:1990, Zaanen:1990,
Anisimov:1991:43}. The Kohn-Sham Hamiltonian is a good starting
point for the kinetic part of the Hamiltonian and can be
conveniently expressed in a basis of linear muffin-tin orbitals
(LMTO's) ~\cite{Andersen:1975}, which need not be orthogonal (see
Appendix~\ref{sec:LMTOH})), as
\begin{eqnarray}
H_{LDA}=\sum_{im,jm',\sigma} %
(\varepsilon_{im}\delta_{im,jm'}+t_{im,jm'})
c_{im\sigma}^{\dagger}c_{jm'\sigma},&&
\label{eq:HLMTO}
\end{eqnarray}
where $i,j$ are atomic site indexes,  $m $ is  orbital one, and
$\sigma$ denotes spin.

It is well known that LDA severely underestimates strong electron
interactions between localized $d$- and $f$-electrons because the
exchange interaction is taken into account only approximately via
the functional of electron density. To correct this situation,
the LDA Hamiltonian can be supplemented with a Coulomb
interaction term between electrons in the localized orbitals
(here we will call them a heavy set of orbitals). The largest
contribution comes from the Coulomb repulsion between electrons
on the same lattice site that we will approximate by the
interaction matrix $U^{i}$ of the heavy shell ($h$) of atom $i$ as%
\begin{equation}
H_{int}={1\over 2}\sum\limits_{ i m\sigma \in h}
   U^{i\sigma\sigma'}_{m_1 m_2 m'_1 m'_2}%
    c^{\dagger}_{i m_1\sigma} c^{\dagger}_{i m_2\sigma'}
    c^{       }_{i m'_2\sigma'} c^{     }_{i m'_1\sigma},
\label{ULMTO}
\end{equation}
where $m$ are orbital and $\sigma $ are spin indexes. In the
diagonal density approximation electron -- electron interaction
matrix $U^{i\sigma\sigma'}_{m_1 m_2 m'_1 m'_2}$  can be
represented using screened Coulomb and exchange vertexes as
\begin{eqnarray}
\label{eq:UJ} %
U_{mm'}&=&\langle mm'|V_{C}|mm'\rangle, \\\nonumber%
J_{mm'}&=&\langle mm'|V_{C}|m'm\rangle,\nonumber
\end{eqnarray}
which are expressed via  Slater integrals $F^{(i)} $, $i=0,2,4,6$
in the standard manner ~\cite{Anisimov:1997:JPCM}. The Slater
integrals can be linked to the average intra--atomic repulsion $U$
and exchange $J$ obtained from, e.g., LSDA supercell procedures
via $U=F^{0}$ and $J=(F^{2}+F^{4})/14$. The ratio $F^{2}/F^{4}$
is to a good accuracy a constant $\sim 0.625$ for $d$-electrons.
Using the Coulomb and exchange matrices we can rewrite the
interaction term as
\begin{eqnarray}\nonumber
H_{int}&=&%
 {1\over 2}\sum\limits_{ i \alpha\alpha'}%
 U^{i}_{\alpha\alpha'}    n_{i \alpha} n_{i \alpha'}%
={1\over 2}\sum\limits_{ i mm'\sigma}%
 U^{i}_{mm'}              n_{i m\sigma} n_{i m' -\sigma}\\%
&       +&{1\over 2}\sum\limits_{ i m\ne m' \sigma}%
(U^{i}_{mm'}-J^{i}_{mm'}) n_{i m\sigma} n_{i m' \sigma},%
\label{eq:Hint}
\end{eqnarray}
where index $\alpha=(m,\sigma)$ combines the orbital and spin
indexes. This equation also provides definition of the
interaction matrix $U_{\alpha\alpha'}$ which will be used further
in the paper.

The LDA Hamiltonian already contains a part of the local
interaction which has to be subtracted to avoid the double
counting. The full Hamiltonian is thus approximated by
\begin{equation}
H = H_{LDA}-H_{dc} +H_{int}= H^0 + H_{int},
\end{equation}
where $H^0$ is the one-particle part of the Hamiltonian and will
play a role of the kinetic term within a DMFT approach. The
double counting correction can not be rigorously derived within
LDA+DMFT. Instead, it is commonly assumed to have a simple static
Hartree-Fock form, just shifting the energies of the heavy set
\begin{equation}
\label{eq:DCT} {H_{dc}}_{\tau m,\tau' m'}(k)=\delta_{\tau m,\tau'
m'}
\delta_{\tau \tau_{h} } E_{dc}. %
\end{equation} %
Here, $\tau$ is atomic index in the elementary unit cell. The
simplest approximation commonly used for $E_{dc}$ is
~\cite{Anisimov:1991,Anisimov:1997}
\begin{equation}
\label{eq:Edc} E_{dc}=U(n_h-{1\over 2}),
\end{equation}
where $n_h=\sum_{m\sigma}n_{m\sigma}$ is the total number of
electrons in the heavy shell (see Appendix~\ref{sec:LMTOH}).

In the spirit of DMFT, the self-energy is assumed to be local,
i.e. $k$-independent, and non-zero only in the block of heavy
orbitals. Therefore it is convenient to partition the Hamiltonian
and the Green's function into the light and heavy set (denoted by
$l$ and $h$, respectively) as
\begin{eqnarray}
\label{eq:Hilbert_real}%
 && G(k,\omega)=  \left[
(\omega+\mu)%
\left( {\begin{array}{*{20}c}
   { O_{hh} } & { O_{hl} }  \\
   { O_{lh} } & { O_{ll} }  \\
\end{array}} \right)_{k}\right . \\\nonumber%
&&\left . %
 - \left( {\begin{array}{*{20}c}
   { H_{hh}^0 } & { H_{hl}^0 }  \\
   { H_{lh}^0 } & { H_{ll} }^0  \\
\end{array}} \right)_k%
-\left( {\begin{array}{*{20}c}
    \Sigma_{hh}(\omega)  & 0  \\
    0      & 0  \\
\end{array}} \right)_{}%
\right]^{-1} ,%
\end{eqnarray}
where $[...]^{-1}$ means matrix inversion, $\mu$ is the chemical
potential and $O$ is the overlap matrix (see Appendix
\ref{sec:no_mbt}).

The main postulate of the Dynamical Mean-Field Theory (DMFT)
~\cite{Georges:1996} formalism is that the self-energy is local,
i.e.~it does not depend on momentum,
$\Sigma(k,\omega)=\Sigma(\omega)$. This postulate can be shown to
be exact in the limit of infinite dimensions provided that the
hopping parameters between different sites are scaled
appropriately. Within this approach, the original lattice problem
can be mapped onto an Anderson impurity model where the local
Green's function and the self-energy, $G_{loc}$ and $\Sigma$, are
identified with the corresponding functions for the impurity
model, i.e.
\begin{equation}
 \Sigma_{imp}(\omega) =  \Sigma(\omega)
\quad\mbox{and}\quad  G_{imp}(\omega) =  G_{loc}(\omega).
\label{SCC}
\end{equation}
Equations (\ref{SCC}) along with the trivial identity
\begin{equation}
\label{eq:localFT}  G_{loc}(\omega) = \sum_{k} G_{}(k,\omega),
\end{equation}
constitute a closed set of self-consistent equations. The only
thing that remains is to solve the Anderson impurity model.

Notice that statement that the self-energy is diagonal is the
basis dependent statement and if $\Sigma (i\omega _{n})$ is
momentum independent in one basis and $U_k$ is a unitary
transformation from one basis to another, and LMTO Hamiltonian,
$H_{LDA}$ in the new basis is given by $U_k H_{LDA}
U^{\dagger}_{k}$, then the self-energy in the new basis $\Sigma '
=U_k \Sigma(i\omega_n) U^{\dagger}_{k}$ is momentum dependent.
Therefore DMFT approximation, if at all valid, is valid in one
basis~\cite{Paul:2002}. Hence, we will work in a very localized
basis where the DMFT approximation is most justified.

In DMFT we construct the self-energy, $\Sigma$, as a solution of
an Anderson impurity model with a non-interacting propagator
(Weiss function) ${\cal G}_{0}$
\begin{eqnarray}
\label{eq:action}
S_{imp} &=& \sum_{\alpha \alpha',\tau \tau ^{\prime } }c_{\alpha }^{+}(\tau )%
 {{\cal G}_{0}}_{\alpha \alpha'}^{-1}(\tau ,\tau ^{\prime})c_{\alpha' }(\tau ^{\prime })%
\\\nonumber%
&+& 
\sum_{\alpha\alpha'\in h}
\frac{U_{\alpha \alpha'}}{2}n_{\alpha }(\tau)n_{\alpha'}(\tau),%
\end{eqnarray}
where $\alpha$ and $\alpha'$ are running over indexes $m\sigma$.
The Weiss function can be linked to the lattice quantities
through the local Green's function and self-energy being are
related to each other by the Dyson equation
\begin{equation}
 G_{loc}(\iomn)^{-1} = {\cal G}_{0}(\iomn)^{-1} -
\Sigma (i\omega _{n}). \label{eq:SE}
\end{equation}
Combining Eq.~(\ref{eq:Hilbert_real}), (\ref{eq:localFT}) and
(\ref{eq:SE}) we finally obtain
\begin{equation}
{\cal G}_{0}^{-1}(i\omega _{n})=\left(\sum_k {1\over
{(i\omega_n+\mu)} O_k -  H^0_k - \Sigma(i\omega
_{n})}\right)^{-1}+\Sigma(i\omega _{n}).%
\label{eq:weiss}
\end{equation}
One can solve a very general impurity model defined by the action
(\ref{eq:action}) and Weiss field (\ref{eq:weiss}). But it is
much cheaper to eliminate the light (weakly interacting) bands
and define an effective action in the subspace of heavy bands
only. In this way, the local problem can be substantially
simplified.

\subsubsection{Downfolding}%

When a group of bands is well separated from the others it is
clear that a reduced description of the problem is possible.  In
the one-electron approach it goes under the name downfolding
~\cite{Andersen:2000}. There are several prescriptions to carry
out this procedure in the context of DMFT. Perhaps, the simplest
approach is to reduce the Hamiltonian ($H^{0}+H_{int}$) (see
Eqs.~(\ref{ULMTO}), (\ref{eq:HLMTO})) to the one having Hubbard
like form for the bands in question. To estimate the hopping
elements one can perform a tight-binding fit. The value of $U$ to
be used is then reduced, since ones computed in the constrained
density functional calculations, it is screened by the bands
which have been eliminated. After this procedure we arrive to a
Hubbard Hamiltonian with a small number of bands
\begin{equation*}
H  =  \sum_{ij m m'} t_{ij}^{m m'} c_{im\sigma}^\dagger
c_{jm'\sigma} +
%
H_{int},
\end{equation*}
where $i,j$ run over lattice sites and $m,m'$ ($\sigma,\sigma'$)
label the orbital (spin) indices of the heavy set of orbitals.

We can also perform the downfolding at the level of DMFT. The
starting point is  Eq.~(\ref{eq:Hilbert_real}) which is used to
get the heavy block Green's function
\begin{equation}
\label{eq:ghh}
G_{hh}(\omega)=\sum_{k}\left[M^k_{hh}-M^k_{hl}{M^k_{ll}}^{-1}M^k_{lh}-\Sigma_{hh}\right]^{-1},
\end{equation}
where the quantity $M^k_{\gamma}$ is defined as
\begin{equation}
M^k_{\gamma}= (\omega+\mu)O_{\gamma}(k)-H_{\gamma}^0(k),
\end{equation}
here $\gamma$ is a double index which is combination of $h$ and
$l$.

The low-energy form of Eq.~(\ref{eq:ghh}) can be found by
expanding around zero frequency and to linear accuracy in
$\omega$ we obtain
\begin{equation}
\label{eq:dmft_downf}
G_{hh}(\omega)=\sum_{k}\left[Z_k^{-1}\omega-\widetilde{H}(k)-\Sigma_{hh}\right]^{-1},
\end{equation}
where renormalization amplitude $Z_{k}$ and effective Hamitonian
are given by
\begin{eqnarray}
Z_{k}^{-1}&=&O_{hh}+K_{hl}K_{ll}^{-1}O_{ll}K_{ll}^{-1}K_{lh}\nonumber\\
&&-O_{hl}K_{ll}^{-1}K_{lh}-K_{hl}K_{ll}^{-1}O_{lh},\nonumber\\
\widetilde{H}(k)&=&H_{hh}^0-K_{hl}K_{ll}^{-1}K_{lh},\nonumber\\
K_{\gamma}&=&H_{\gamma}^0-\mu O_{\gamma}. \label{eq:h_eff0}
\end{eqnarray}
Further, we can choose a new base in the heavy block such that the
local Green's function has the usual form. First we define the
average renormalization amplitude
\begin{equation}
Z_{\alpha\beta}=\sum_k {Z_{k}}_{\alpha\beta},
\end{equation}
and choose the transformation matrix such that
\begin{equation}
S^{\dagger}Z^{-1}S = 1.
\end{equation}
The new overlap matrix and effective Hamiltonian become
\begin{eqnarray}
O_{eff}(k) &=& S^{\dagger} Z_k^{-1} S\;,\\
H_{eff}(k) &=& S^{\dagger} \widetilde{H}(k) S + \mu O_{eff}(k)\label{eq:h_eff}\;,\\
\Sigma &=& S^{\dagger} \Sigma_{hh} S \label{eq:sig_upf} \;,
\end{eqnarray}
and finally the local Green's function in the new base takes the
form
\begin{equation}
\label{eq:RD} G_{hh} =
\sum_k\left[(\omega+\mu)O_{eff}(k)-H_{eff}(k)-\Sigma\right]^{-1},
\end{equation}
with the Dyson equation
\begin{equation}
G_{hh}^{-1} = {\cal G}_{0hh}^{-1} - \Sigma\; , %
\label{eq:ghh1}
\end{equation}
and corresponding Weiss function defined as
\begin{equation}
{\cal G}_{0hh} = \left[(\omega+\mu)-\Delta \right]^{-1},
\end{equation}
where hybridization function $\Delta$ regularly behaves at
infinity.

The self-energy of the reduced model (\ref{eq:RD}) is local,
therefore the DMFT treatment is applicable. In this case, only the
heavy bands need to be considered in calculation which greatly
simplifies the complexity of the problem.

The reduced model, however, has in general more complicated
Coulomb interaction matrix than we chose in the original model.
If we assume, that the diagonal components of $Z$ are much larger
than the off-diagonal, we obtain the same simple Hubbard-type
interaction term within reduced model for the heavy block. The
Coulomb interaction is however screened by the light bands and is
reduced to
\begin{equation}
U_{\alpha\beta}\rightarrow U_{\alpha\beta} Z_{\alpha}Z_{\beta}.
\end{equation}
We can estimate the magnitude of the reduction of Coulomb
repulsion by evaluating the above quantity explicitly. In the
case the original base is orthogonal, i.e. $ O=1$, we get
\begin{equation}
{\Delta U\over U} \sim -{1\over N}{\rm
Tr}\left[H^0_{hl}\left(1\over H^0_{ll}-\mu\right)^2
H^0_{lh}\right], \label{eq:delta_u}
\end{equation}
where $N$ is the number of heavy bands.

In the case when the reduced model corresponding to
Eq.~(\ref{eq:RD}) is degenerate and the basis is orthogonal (the
overlap matrix $O_{eff}(k)$ is equal to unity) the self-energy
and local Green's function are also degenerate and diagonal. Then
the momentum sum in Eq.~(\ref{eq:RD}) can be replaced by the
integral over energy and the local Green's function can be
calculated using the standard Hilbert transformation
\begin{equation}
G^{}(i\omega _{n})=\int_{-\infty }^{+\infty }d\varepsilon \,{\frac{{%
D(\varepsilon )}}{{i\omega _{n}+\mu -\Sigma (i\omega
_{n})-\varepsilon }}}. %
\label{eq:Hilbert}
\end{equation}
Here, the density of states $D(\varepsilon)$ is the density of
states of the kinetic part of the reduced model $H_{eff}(k)$ in
Eq.~(\ref{eq:RD}).

Other groups have used for $D(\varepsilon )$ rescaled partial DOS
as discussed in Appendix~\ref{sec:LMTOH}. Possibility to use the
Hilbert transformation is substantially simplifies the
calculation procedure and brings a number of conceptual
simplifications ~\cite{Held:2001}.

\subsubsection{Upfolding}%

Upfolding is a procedure which is ``inverse" to the downfolding
one. One needs to use Eq.~(\ref{eq:sig_upf}) to transform the
self-energy obtained from DMFT calculations, $\Sigma$, to the
one, $\Sigma_{hh}$, which is inserted to the original LDA
Hamiltonian in order to compute the local Green's function (GF)
$G_{}(i\omega _{n})$. The local GF with upfolded self-energy reads
\begin{eqnarray}
\label{eq:Hilbert_real_up}%
 && G_{}(i\omega _{n})=  \int dk \left[
(i\omega_{n}+\mu)%
\left( {\begin{array}{*{20}c}
   { O_{hh} } & { O_{hl} }  \\
   { O_{lh} } & { O_{ll} }  \\
\end{array}} \right)_{k}\right . \\\nonumber%
&&\left . %
 - \left( {\begin{array}{*{20}c}
   { H_{hh} } & { H_{hl} }  \\
   { H_{lh} } & { H_{ll} }  \\
\end{array}} \right)_k%
-\left( {\begin{array}{*{20}c}
    \Sigma_{hh} -  H_{dc}& 0  \\
    0      & 0  \\
\end{array}} \right)_{(i\omega _{n})}%
\right]^{-1} ,%
\label{eq:gf_upfolded}
\end{eqnarray}
where $\mu$ is the LDA chemical potential and $ H_{dc}$ is the
double counting term Eq.~(\ref{eq:DCT}). Instead of using formula
(\ref{eq:Edc}), we rather deduce the constant shift of the heavy
set of bands by equating the total number of electrons to the
integral of the spectral function
$$
A(\omega ) = -\frac{1}{\pi}{\rm Im}\sum\limits_k {}
\sum\limits_{\alpha\beta} {} G_{\alpha\beta}^{} (k,\omega
)O_{\alpha\beta}^k,
$$
multiplied by the Fermi function.

\subsubsection{Algorithm to solve DMFT equations}%

To close the set of DMFT equations, a method to solve the local
problem is required. In the following, we will focus our attention
on two impurity solvers: QMC and SUNCA. In section
\ref{sec:imsolvers}, we will briefly review both methods while a
detailed comparison between the results obtained by those two
approaches is given in section \ref{sec:comp_qmc_sunca}. Bellow,
we summarize basic steps in the DMFT self-consistent scheme that
delivers the local self-energy - crucial quantity to calculate
transport and optical properties of a solid.

We started the iteration by a guess for the Weiss field ${\cal
G}_0^{-1}$ from which the local Green's function $G_{loc}$ was
calculated by one of the impurity solvers. The self-energy was
then obtained by the use of the Dyson equation ~(\ref{eq:ghh1}).
Momentum summation over the Brillouin zone, Eq. ~(\ref{eq:RD}),
delivers a new guess for the local Green's function and through
the Dyson equation also for the Weiss field ${\cal G}_0^{-1}$.
The iteration is continued until the convergence is found to the
desired level. The scheme can be illustrated by the following
flow-chart
\[
{\cal G}_0^{-1}\stackrel{\hbox{\small\it IMP
solver}}{\longrightarrow } {G} \stackrel{\hbox{\small\it
DE}}{\longrightarrow } {\Sigma}\;\stackrel{\hbox{\small\it DMFT
SCC}}{\longrightarrow }{\cal G}_0^{-1},
\]
where ``DE" stands for the Dyson equation ~(\ref{eq:ghh1}) and
``DMFT SCC" means the DMFT self-consistent condition Eq.
 ~(\ref{eq:RD}) or ~(\ref{eq:Hilbert}).

The QMC impurity solver is defined in imaginary time, $\tau$,
therefore the following additional Fourier transformations
between imaginary time and Matsubara frequency points are
necessary
\[
{\cal G}_0(i\omega)\stackrel{\hbox{\small\it
IFT}}{\longrightarrow } {\cal
G}_0(\tau)\;\stackrel{\hbox{\small\it  QMC}}{\longrightarrow } {
G(\tau)}\;\stackrel{\hbox{\small\it FT}}{\longrightarrow }{
G(i\omega)}.
\]
Here FT and IFT are Fourier and inverse Fourier transformations,
respectively. After the self-consistency is reached, the analytic
continuation is required to obtain the real-frequency
self-energy. This issue is addressed in section
\ref{sec:acontinuation}.

The SUNCA method is implemented on real frequency axis to avoid
the ill-posed problem of analytic continuation. As an input, it
requires the bath spectral function $A_c(\omega) =
-{1\over\pi}{\rm Im}{\cal G}_0^{-1}(\omega)$ and delivers the
local spectral function $A_d(\omega)=-{1\over\pi}{\rm
Im}G(\omega)$
\[
{ A_c(\omega)}\stackrel{\hbox{\small\it SUNCA}}{\longrightarrow }
{ A(\omega)}\stackrel{\hbox{\small\it KK}}{\longrightarrow }{
G(\omega)}.
\]
The real part of the local Green's function is obtained by the
use of the Kramers-Kronig relation (KK).

\section{Impurity solvers} %
\label{sec:imsolvers}

Among many methods used to solve the impurity problem we chose
the Quantum Monte Carlo  method~\cite{Hirsch:1986} and
symmetrized finite-U NCA method~\cite{Haule:2001}. In this section
we briefly describe both of them.

\subsection{The Quantum Monte Carlo method}%
\label{sec:qmc}

There are well known advantages and disadvantages of the QMC
method and our choice is spurred by the fact that despite being
slower than other methods the QMC is well controlled, exact
method.  As an input the QMC procedure gets Weiss function ${\cal
G}_{0}(\tau)$ and as an output it produces Green's function
$G(\tau )$. We remind reader  major steps taken for the QMC
procedure. Usually one starts with impurity effective action $S$
\begin{eqnarray}
S_{eff}&=& - \int^{\beta}_{0} d\tau d\tau' \sum_{\alpha}
c^{+}_{\alpha}(\tau) {{\cal G}_{0}}_{\alpha}^{-1}(\tau,\tau')
c_{\alpha}(\tau^{\prime})  \\\nonumber
&+&\frac{1}{2}\,\int^{\beta}_{0}d\tau\, \sum_{\alpha ,\alpha '}
U_{\alpha\alpha'}n_{\alpha}(\tau)n_{\alpha'}(\tau),
 \label{eq:simp}
\end{eqnarray}
where $\{c,\ c^+\}$ operators are fermionic annihilation and
creation operators of the lattice problem, $\alpha =\{m,\sigma\}$.

The first what we should do with the action (\ref{eq:simp})  is
to discretize it in imaginary time space with time step $\Delta
\tau$ such that $\beta=L\Delta\tau$, and $L$ is the number of
time intervals
\begin{eqnarray}
S_{eff} &\rightarrow& \sum_{\alpha ,\tau \tau ^{\prime }
}c_{\alpha }^{+}(\tau ) {{\cal G}_{0}}_{\alpha }^{-1}(\tau ,\tau
^{\prime })c_{\alpha  }(\tau ^{\prime }) \\\nonumber &+&
\frac{1}{2}\sum_{\alpha ,\alpha '}U_{\alpha\alpha'} n_{\alpha
}(\tau)n_{\alpha '}(\tau). \label{eq:simdiscret}
\end{eqnarray}
The next step is to get rid of the interaction term $U$ by
substituting it by summation over Ising-like auxiliary fields.
The decoupling procedure is called the Hubbard-Stratonovich
transformation~\cite{Hirsch:1983,Takegahara:1992}
\begin{equation}
\exp \{  - \Delta \tau  \{U_{\alpha\alpha'} n_{\alpha}  n_{\alpha
'} -
\frac{1}{2}(n_{\alpha}   + n_{\alpha '} )\} \}  =%
\label{eq:HStransformation}
\end{equation}
$$
\frac{1}{2}\sum\limits_{S\alpha \alpha ' =  \pm 1} {} \exp \{
\lambda_{\alpha \alpha '}
S_{\alpha \alpha '} (n_\alpha   - n_{\alpha '} )\},  $$%
where $ \cosh \lambda_{\alpha \alpha '}  = \exp (\frac{{\Delta
\tau U_{\alpha \alpha '} }}{2}) $, $S_{\alpha \alpha '}(\tau_l)$
are auxiliary Ising fields at each time slice.

In the one-band Anderson impurity model  we have only one
auxiliary Ising field $S(\tau_l)=\pm 1$ at {\sl each time slice},
whereas in the multiorbital case  number of auxiliary
fields is equal to  number of $\alpha, \alpha'$ pairs, i.e. ${}^{\alpha}C_2$. %
Applying the Hubbard-Stratonovich transformation at each time
slice we bring the action to the quadratic form with the
partition function
\begin{equation}
 Z = Tr_{\{S_{\alpha \alpha '}(\tau)\}} \prod_{\alpha } \det
G^{-1}_{\alpha ,\{S_{\alpha \alpha '}(\tau)\} }\; ,
\end{equation}
where  GF in terms of auxiliary fields  $G^{-1}_\alpha $  reads as
\begin{equation}
G^{-1}_{\alpha ,\{S_{\alpha \alpha '} \}}(\tau, \tau') = {{\cal
G}_0}^{-1}_{\alpha }(\tau, \tau')e^{V} -
(e^{V}-1)\delta_{\tau,\tau'},
\end{equation}
with interaction matrix
\begin{equation}
 V_{\tau}^{\alpha }  = \sum\limits_{\alpha '( \ne \alpha )} \lambda_{\alpha \alpha '}
 {S_{\alpha \alpha '} (} \tau ) \sigma_{\alpha \alpha '},
\end{equation}%
where\\
 $ \sigma_{\alpha \alpha '}  =  + 1 $ for $\alpha  < \alpha ' $\\
 $ \sigma_{\alpha \alpha '}  =  - 1 $ for   $\alpha  > \alpha '$.\\ %

Once the quadratic form is obtained one can apply Wick's theorem
at each time slice and  make the Gaussian integration by Grassmann
variables to get the full interacting GF
\begin{eqnarray}
\label{eq:gfull}
 G_{\alpha }(\tau,\tau') &=& {\frac{{1}}{{Z}}}\,
Tr_{\{ S_{\alpha \alpha '} \} } \\\nonumber
 & &
G_{\alpha ,\{ S_{\alpha \alpha '}\}}(\tau,\tau')\prod_{\alpha '}
\det G^{-1}_{\alpha ',\{S_{\alpha \alpha '}\}}.%
\end{eqnarray}

To evaluate summation ~Eq.~(\ref{eq:gfull}) one uses Monte Carlo
stochastic sampling. The product of determinants is interpreted
as the stochastic weight and auxiliary spin configurations are
generated by a Markov process with probability proportional to
their statistical weight. More rigorous derivation can be find
elsewhere ~\cite{Georges:1996}, \cite{Takegahara:1992}.

Since the QMC method produces results in complex time
($G(\tau_m)$ with $\tau_m=m\Delta\tau$, $m=1...L$) and the DMFT
self-consistency equations make use of the frequency dependent
Green's functions and self-energies we must have an accurate
method to compute Fourier transforms from the time to frequency
domain. This is done by representing the functions in the time
domain by a cubic splined functions which should go through
original points with condition of continuous second derivatives
imposed. Once we know cubic spline coefficients we can compute the
Fourier transformation of the splined functions analytically (see
Appendices~\ref{app:spline} and \ref{app:moments}).

\subsection{The SUNCA}%
\label{sec:sunca}

In this section we briefly review the second method, used to solve
the multiorbital Anderson impurity model, called Symmetrized
finite-U NCA method \cite{Haule:2001}. The method is based on the
self-consistent perturbation theory with respect to the
hybridization strength between the effective bath and the local
system and is therefore exact in the atomic limit. However, this
method sums up an infinite class of skeleton diagrams and takes
into account a subclass of singular vertex corrections that are
necessary to obtain the correct dynamic low-energy Fermi-liquid
scale \cite{Haule:2001} and correct position of the
Abrikosov-Suhl resonance. The SUNCA approximation does not
contain only non-crossing diagrams but rather all the three-point
vertex corrections of ladder-type and should not be confused with
the usual finite-U non-crossing approximation.

The SUNCA approach presents the advantage to provide directly real
frequencies one-particle Green's functions. Local quantities like
densities of states can therefore be computed for any regime of
parameters without having to perform analytical continuation.
Furthermore, the SUNCA method can be applied to arbitrary
multiband degenerate Anderson impurity model with no additional
numerical cost. This is an important advantage compared to some
other methods like Quantum Monte Carlo or exact diagonalization.
The method is especially relevant for systems with large orbital
degeneracy such as systems with $f$-electrons.

The pathologies that severely limit the usefulness of the
non-crossing approximation in the context of DMFT are greatly
reduced with inclusion of ladder-type vertex corrections.
Nevertheless, they do not completely remove the spurious peak
that forms at temperatures substantially below the Kondo
temperature. To overcome this shortcoming, we employed an
approximate scheme to smoothly continue the solution down to zero
temperature. This was possible because at the break-down
temperature the solution of SUNCA equations shows an onset of the
Fermi-liquid state. As we will show in the subsequent chapters by
comparison with QMC, SUNCA gives correct quasiparticle
renormalization amplitude $Z$ and the real part of the
self-energy at zero frequency approaches the Luttinger value. The
imaginary part of the self-energy, however, has a narrow spurious
deep on top of the parabola that is formed around zero
frequency.  To remove the deep, we matched the Fermi-liquid
parabolic form for imaginary part of the self-energy in the
small-window of the deep such that it smoothly connected the
intermediate frequency region where the parabola was formed. We
numerically found that this SUNCA pathology is rapidly reduced
with increasing the number of bands i.e. it is much less severe in
the case of three-band model than in one-band case.

Next, we give some details of the auxiliary particle technique
together with the definition of the SUNCA approximation. Within
the auxiliary diagrammatic method, the local degrees of freedom
can be represented by auxiliary particles. To each eigenstate of
the local Hamiltonian $H_{loc}\left|n\right> = E_n\left|n\right>$
we assign an auxiliary operator $a_n$ such that $\left|n\right> =
a_n^\dagger\left|vac\right>$. A general Anderson impurity model
can then be expressed by
\begin{eqnarray}
H=\sum_n E_n a_n^\dagger a_n + \sum_{k,m}
\varepsilon_{km}c_{km}^\dagger c_{km} + \nonumber\\
\sum_{k,m,n,n^\prime}(V_{km}^* F^{m}_{n,n^\prime} a_n^\dagger
a_{n^\prime} c_{km}+h.c.), \label{eq:H_AIM}
\end{eqnarray}
where
$F^{m}_{n,n^\prime}=\left<n\right|d_m^\dagger\left|n^\prime\right>$,
$d_m^\dagger$ is a creation operator for the local electron and
$c_{km}^\dagger$ creates an electron in the bath and $m$ stands
for the spin and band index. In order that electrons are
faithfully represented by the auxiliary particles, two conditions
must be satisfied: \vskip 0.2cm \noindent (i) An auxiliary
particle $a_n$ must be boson (fermion) if the state
$\left|n\right>$ contains even (odd) number of electrons. \vskip
.2cm \noindent (ii) The local charge $Q \equiv \sum_n
a_n^{\dagger}a_n$ must be equal to one at all times $Q=1$
expressing the completeness relation for the local states $\sum_n
\left|n\right>\left<n\right|=1$. \vskip 0.2cm \noindent The first
condition merely request some care that has to be taken in
evaluating diagrams while the second constraint, projection onto
the physical Hilbert space, is somewhat more involved but can
still be done exactly. The term $\lambda Q$ can be added to the
Hamiltonian and the limit $\lambda\rightarrow\infty$ has to be
taken after the analytic continuation to the real frequency axes
is performed. Taking this limit, actually leads to a substantial
simplification of the analytic continuation. Namely, when
evaluating self-energies for the pseudo-particles, only the
integrals around the branch-cuts of the bath electron Green's
function have to be considered while the integrals around the
auxiliary particle Green's functions vanish by the projection.

The physical local Green's function (electron Green's function in
Q=1 subspace) can eventually be calculated with the help of the
Abrikosov trick \cite{Abrikosov:1965} which states that the
average of any local operator that vanishes in the $Q=0$ subspace
is proportional to the grand-canonical (all $Q$ values allowed)
average of the same operator
\begin{equation}
\langle A \rangle_{Q=1} = \lim_{\lambda\rightarrow\infty}{\langle
A\rangle_G \over \langle Q\rangle_G}.
\end{equation}
By realizing that the local Green's function is proportional to
the bath electron T-matrix, we have
\begin{equation}
G_{loc} = \lim_{\lambda\rightarrow\infty}{1\over V^2 \langle Q
\rangle_G} \Sigma_c, \label{eq:locG}
\end{equation}
where $\Sigma_c$ is the bath electron self-energy calculated in
the grand-canonical ensemble.

In the case of degenerate Anderson impurity model, an important
simplification occurs in the Hamiltonian Eq.~(\ref{eq:H_AIM}).
Namely, pseudo-particles, corresponding to the states with the
same number of electrons on the impurity site, are degenerate
with energies $E_M = -M\mu + U M(M-1)/2$, and pseudo Green's
functions
\begin{equation}
G_M(i\omega) = 1/(i\omega-\lambda-E_M-\Sigma_M(i\omega)),
\label{eq:greens}
\end{equation}
where $M$ is the number of electrons on the impurity. In the case
of Anderson impurity model with $N/2$ bands, we need to consider
only $N+1$ different propagators instead of dealing with $2^N$
pseudo-particles. Furthermore, if $U$ is large and we are
interested in doping levels not too far from an integer filling
with $M$ electrons, i.e. close to the Mott-insulating state, it
is reasonable to assume that only fluctuations between local
states with $M-1$, $M$ and $M+1$ electrons on the impurity need
to be considered. The single particle spectra will thus consist
of lower Hubbard band, upper Hubbard band and a quasiparticle
resonance while we ignore other Hubbard bands that are even more
far away from the chemical potential. For example, in the case
the filling is less than $1.5$ we take into account empty state,
singly occupied and doubly occupied states on the impurity while
the triple occupancy is neglected. This is equivalent to adding
to the original Hamiltonian an infinite three-particle
interaction.

\begin{figure}
\begin{center}
\includegraphics[width=3.2in,angle=0]{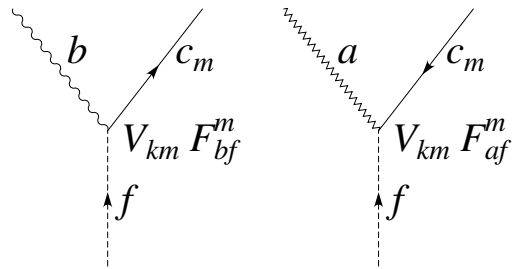}
\end{center}
\caption{ \label{fig:vertex} The two bare vertices considered in
the case of degenerate Anderson impurity model.
Throughout this paper, conduction-electron $c$ propagator is
represented by solid line, while wiggly, dashed and zigzag lines
correspond to the pseudo-particles with $M-1$, $M$ and $M+1$
electrons on the impurity, respectively.
}
\end{figure}
In the following, we will use letter $b$ to denote the auxiliary
particles with $M-1$ electrons on the impurity, $f$ for $M$ and
$a$ for $M+1$ electrons on the local level, respectively. In
Fig.~\ref{fig:vertex} we show the two bare three-point vertices
that are left to us in the case of degenerate Anderson impurity
model when considering those three types of states. Note that
pseudo-particle $b$ is $\left(\begin{array}{c}N\\
M-1\end{array}\right)$ degenerated, $f$ is
$\left(\begin{array}{c}N\\ M\end{array}\right)$ and $a$ is
$\left(\begin{array}{c}N\\ M+1\end{array}\right)$ degenerated,
respectively. In case $M$ is odd (even), particles $a$ and $b$ are
bosons (fermions) while $f$ are fermions (bosons).

\begin{figure}
\begin{center}
\includegraphics[height=3.2in,angle=-90]{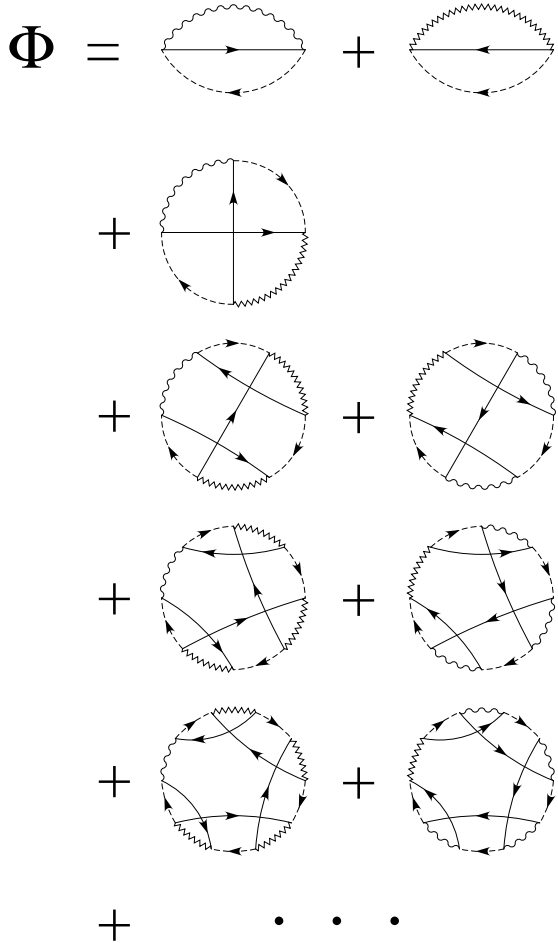}
\end{center}
\caption{ \label{fig:LuttWard} Diagrammatic representation of the
SUNCA generating functional to describe the degenerate Anderson
impurity model. }
\end{figure}
The SUNCA approximation is a conserving approximation defined by a
Luttinger-Ward type functional $\Phi$ from which all
self-energies are obtained as a functional derivatives,
$\Sigma_a={\delta\Phi\over\delta G_a}$. The building blocks of
$\Phi$ are dressed Green's functions of pseudo-particles $G_b$
(depicted as a wiggly line), $G_f$ (dashed line), $G_a$ (zigzag
line) and bath electron $G_c$ (solid line). Due to the exact
projection, only pseudo-particles are fully dressed while bath
electron Green's function is not dressed.

The choice of diagrams was motivated by the Schrieffer-Wolf
transformation showing that both fluctuations from $M$ to $M-1$
and $M$ to $M+1$ have to be considered totally symmetrically. In
the case of one-band model we have for the exchange coupling
$J=J_1+J_2=V^2({{1\over\mu}+{1\over U-\mu}})$. The exact Kondo
temperature is proportional to
$\exp(-\pi/J)=\exp(-\pi/(J_1+J_2))$. It is well known that if one
takes into account only non-crossing diagrams, the Kondo scale of
the resulting approximation steeply drops with decreasing $U$ and
is of the order of $\exp(-\pi/J_1-\pi/J_2)$ which can be orders
of magnitude wrong. This happens because the simultaneous
fluctuations between all three types of pseudo-particles are
neglected. As was shown in Ref.~\onlinecite{Haule:2001}, one needs
to sum up an infinite number of skeleton diagrams (SUNCA
diagrams) to recover the correct exchange coupling consisting of
two terms $J_1$ and $J_2$, generated from the two types of
vertices depicted in Fig.~\ref{fig:vertex}.

The SUNCA Luttinger-Ward functional, shown in
Fig.~\ref{fig:LuttWard}, consists, in addition to NCA
contributions (first two diagrams), of the diagrams where two
conduction electron lines cross only ones and the rest of the
electron lines cross twice. Diagrams not included in SUNCA have
higher order crossings. The lowest order neglected are of
CTMA-type\cite{Costi:1994,Kroha:2003} where all conduction
electrons cross exactly twice. Note that due to the projection,
any contribution to the Luttinger-Ward functional consists of a
single ring of pseudo-particles since at any moment in time there
must be exactly one pseudo-particle in the system.
\begin{figure}
\begin{center}
\includegraphics[height=3.2in,angle=-90]{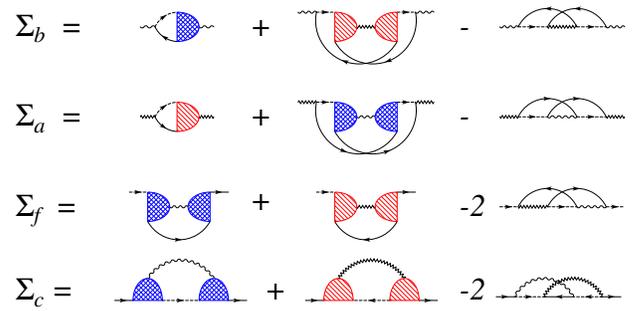}
\end{center}
\caption{ \label{fig:selfen}
Diagrammatic representation of the self-energies, derived from
SUNCA Luttinger-Ward functional (Fig.~\ref{fig:tmatrix}) in terms
of the renormalized hybridization vertices, defined in
Fig.~\ref{fig:tmatrix}.  In each line the third diagram is
subtracted in order to avoid double counting of terms within the
first two diagrams.
}
\end{figure}
\begin{figure}
\begin{center}
\includegraphics[height=3.2in,angle=-90]{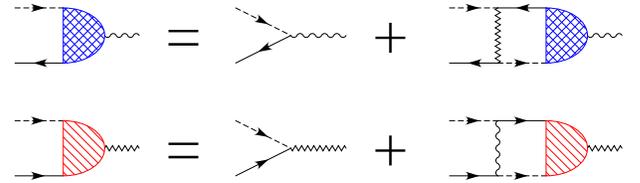}
\end{center}
\caption{ \label{fig:tmatrix} Diagrammatic representation of the
Bethe-Salpeter equations for ladder-type three-point vertex
function. }
\end{figure}
The SUNCA self-energies, obtained by differentiating the
Luttinger-Ward functional, are shown in Fig.~\ref{fig:selfen}. The
part of the pseudo-particle ring, where conduction electrons cross
exactly twice, can be rewritten in the ladder type T-matrix.  The
additional two conduction lines, that cross only once, enable us
to close T-matrix and convert it into three-point vertex shown in
Fig.~\ref{fig:tmatrix}. This is important from practical point of
view since the tree-point vertex is numerically much more
tractable than the full four-point vertex.

The expressions for the self-energies $\Sigma_{\mu}$ defined by
Fig.~\ref{fig:selfen} and Fig.~\ref{fig:tmatrix}, together with
the definitions of the Green's functions, Eq.~(\ref{eq:greens}),
constitute a set of non-linear integral equations for
$\Sigma_{\mu}, \mu=a,b,f$. The local Green's function is
calculated from the Eq.~(\ref{eq:locG}) using the
self-consistently determined auxiliary propagators $G_{\mu}$. The
SUNCA equations are given explicitly in
Appendix~\ref{app:SUNCA_EQ}.

The SUNCA equations have been evaluated numerically by iteration.
The two self-consistent loops, SUNCA and DMFT, can be merged
together into one single set of equations. Starting with the
initial guess for the bath spectral function $A_c$ and
pseudo-particle Green's functions $G_{\mu}$, the first guess to
the T-matrix defined in Fig.~\ref{fig:tmatrix} is determined.
With this, the pseudo-particle self-energies as well as the local
Green's function may be deduced. From the DMFT self-consistent
condition, the new bath spectral-function is calculated. With the
updated pseudo-particle Green's functions and the new bath
spectral-function one determines next approximation for the
T-matrix, pseudo-particle self-energies and local Green's
function. The iteration is continued, until the convergence is
found to the desired level.

\subsection{Analytic continuation of the self-energy}%
\label{sec:acontinuation}

The QMC simulation produces Green's function $G(\tau)$ of
imaginary time $\tau=i t$ or equivalently Green's function and
the self-energy defined in Matsubara frequency points. However,
real-frequency self-energy is needed to obtain transport
quantities. The analytic continuation of QMC data is required,
which is an ill-posed problem and altogether hopeless if the
precision of data is not extremely good and if the statistical
errors are not taken into account properly. As is well known,
P\'ade method is not very useful for analytic continuation of
noisy QMC data. The maximum entropy method (MEM)
\cite{Jarrell:1996} tries to overcome this problem by adding an
entropy term to the functional to be minimized. This is one of
the best methods present available and usually produces
real-frequency Green's function of relatively high quality
provided the data are carefully analyzed. We refer the reader to
the original literature for the details \cite{Jarrell:1996}.

However, the quasiparticle peak for realistic density of states
can have quite reach structure since at low temperature it tries
to reproduce the LDA bands around the Fermi-level, i.e., the
spectral function approaches the LDA density of states contracted
for the quasiparticle renormalization amplitude $Z$, $A(\omega) =
\rho(\omega/Z+\mu_0)$. The maximum entropy method has a tendency
to smear out this reach structure because of the entropy term. At
low temperature, this can lead to overshooting of spectral
function and subsequently to the non-physical self-energy that
ruins the causality. To avoid this pathology, we sometimes found
useful to directly decompose the singular kernel with the
Singular Value Decomposition (SVD). When constructing the real
frequency data, we took into account only those singular values,
which are larger than precision of the QMC data.

Imaginary time Green's $G(\tau)$ can be expressed by the spectral
function as
\begin{equation}
G(\tau) = -\int d\omega f(-\omega)e^{-\tau\omega} A(\omega)\ ,
\end{equation}
or in discretized form
\begin{equation}
G_\tau = -\sum_\omega f(-\omega) e^{-\tau\omega} A_\omega
\Delta_{\omega} = \sum_{\omega m} V_{\tau m} S_m U_{m\omega}
A_\omega, \label{eq:SVD}
\end{equation}
where $U U^\dagger=1$ and $V^\dagger V=1$ are orthogonal matrices
and $S$ is diagonal matrix of singular values. The inversion is
than simply given by
\begin{equation}
A_\omega = \sum_{m,\tau} U_{m\omega} {1\over S_m} V_{\tau m}
G_\tau . \label{eq:SVDi}
\end{equation}
The magnitude of singular values drops very fast and only first
few terms in the upper sum can be determined from the QMC data.
The rest of the information, that determines mostly higher
frequency points, can be acquired from the SUNCA spectral
function. We therefore approximated the sum in
Eq.~(\ref{eq:SVDi}) by
\begin{eqnarray}
&& A_\omega = \sum_{m\le M,\tau}
U_{m\omega}\;\alpha_m^{QMC}+\sum_{m>M}
U_{m\omega}\;\alpha_m^{SUNCA} \ ,\nonumber \\
&& \alpha_m^{QMC} = \sum_\tau {1\over S_m}V_{\tau m} G_\tau\nonumber \ , \\
&& \alpha_m^{SUNCA} = \sum_{\omega^\prime} U_{m\omega^\prime}
A^{SUNCA}_{\omega^\prime}, \label{eq:SVDsum}
\end{eqnarray}
where $M$ can be determined by the precision of the QMC data,
i.e., $\sum_{\tau} V_{\tau M}\delta G_\tau >S_M$.

\begin{figure}[h]
\begin{center}
\includegraphics[width=3.2in]{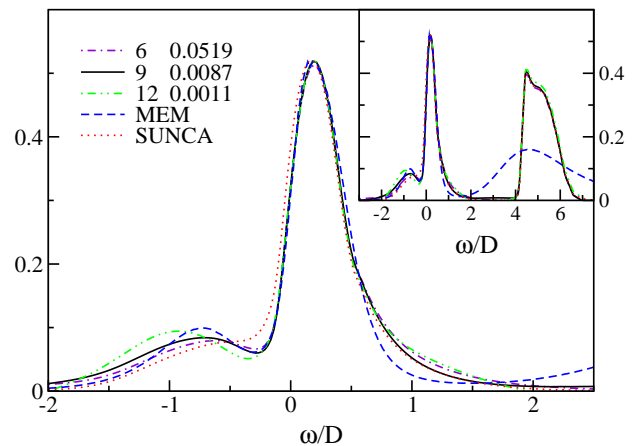}
\end{center}
\caption{ \label{fig:SVD} Spectral function for semicircular DOS,
inverse temperature $\beta=16$ and density $n=0.8$. Dot-dashed,
full and double-dot-dashed curve correspond to the sum
(~\ref{eq:SVDsum}) with $M$ chosen to be $6$, $9$ and $12$,
respectively.  In legend, we also print the lowest singular value
taken into account ($S_M$).  For comparison we show maximum
entropy spectra (dashed curve) and SUNCA spectra (dotted line).
The inset shows the same spectra in a broader window.  }
\end{figure}

We plot the sum (\ref{eq:SVDsum}) in Fig.~\ref{fig:SVD} where
first $3$, $6$ or $9$ coefficients were obtained from the QMC
data.  The corresponding smallest singular value is printed in
the legend of the same figure.  For comparison, we also display
the spectral function obtained by the maximum entropy method and
the SUNCA solution for the same parameters. The difference
between the various curves gives as a rough estimate for the
accuracy of the technique. As we see, the quasiparticle resonance
is obtained by reasonably high accuracy, while the Hubbard band
is  determined with less accuracy.  In the inset of
Fig.~\ref{fig:SVD} we plot the same curves in a broader window.
As we see, the SVD does not guarantee the spectra to be positive
at higher frequencies. This however does not prevent us to
accurately determine most of the physical quantities.

\begin{figure}[h]
\begin{center}
\includegraphics[width=3.2in]{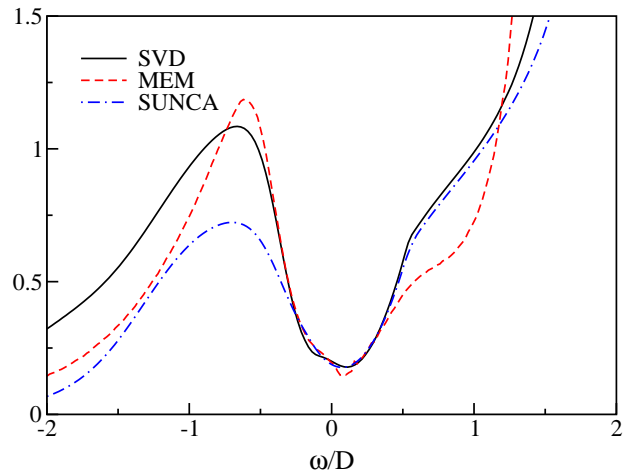}
\end{center}
\caption{ \label{fig:SVD1} Imaginary part of the self-energy
obtained from the Green's function by the inverse of the Hilbert
transform. Full-line was obtained by the Singular Value
decomposition, dashed by the maximum entropy method and
dot-dashed by SUNCA. Parameters used are the same as in
Fig.~\ref{fig:SVD} }
\end{figure}

Within DMFT, the real frequency self-energy can be obtained from
the local Green's function by the inversion of the Hilbert
transform. Although the implementation is very straightforward, we
will briefly mention the algorithm we used. In the high-frequency
regime, we can expand the Hilbert transform in terms of moments
of the DOS as
\begin{equation}
w(z) = \int {D(\varepsilon) d\epsilon\over z-\varepsilon} =
\sum_n {\langle\varepsilon^n\rangle\over z^{n+1}}.
\end{equation}
The series can be inverted and solved for $z$
\begin{eqnarray}
z(w) &=& {1\over w} +\langle\varepsilon\rangle +
(\langle\varepsilon^2\rangle-\langle\varepsilon\rangle^2) w \nonumber\\
&+&
(\langle\varepsilon^3\rangle-3\langle\varepsilon^2\rangle\langle\varepsilon\rangle+
2\langle\varepsilon\rangle^3) w^2 + ... \ . \label{eq:expan}
\end{eqnarray}
For most of the frequency points, the expansion up to some higher
power ($\sim w^8$) gives already an accurate estimation for the
inverse function. However, when $w$ gets large, we need to use
one of the standard root-finding methods to accurately determine
the solution. This is however much easier than general
root-finding in complex plane since we always have a good
starting guess for the solution.  We start evaluating the inverse
function at high frequency where the absolute value of $G$ is
small and we can use the expansion in Eq.~(\ref{eq:expan}). Then
we use the fact that Green's function is a continuous function of
a real frequency and we can follow the solution from frequency
point to frequency point by improving it with few steps of a
secant (or Newton) method. A special attention, however, must be
paid not to cross the branch-cut and get lost in the non-physical
complex plane. Therefore, each secant or Newton step has to be
shortened if necessary.  The self-energy is finally expressed by
the inverse of Hilbert transform $w^{-1}$ as
\begin{equation}
\Sigma = \omega+\mu-w^{-1}(G).
\end{equation}
 Fig.~\ref{fig:SVD1} shows the imaginary part of
self-energy obtained by both analytic-continuation methods. As a
reference and comparison we also show the results obtained by
SUNCA method, which is defined and evaluated on real frequency
axes and hence does not require analytic continuation. The
low-frequency part of the self-energy is again very reliably
determined and does not differ for more than 3\%.

\section{Transport Computation}%
\label{sec:transport_theory}

\subsection{Transport theory} 
\label{subsec:transpot}

The transport parameters of the system are expressed in terms of
so called kinetic coefficients, denoted here by $A_{m}$.  The
equation for the electrical resistivity, $\rho$, is given by
\begin{equation}
\label{eq:resistivity}
\rho = \frac{\kb T}{e^{2}}\frac{1}{A_{0}},
\end{equation}
and the thermopower, $S$, and the thermal conductivity, $\kappa$,
are expressed through
\begin{equation}
  \label{eq:thermalparameters}
  S = -\frac{\kb}{|e|}\frac{A_{1}}{A_{0}},
  \qquad
  \kappa = \kb\left(A_{2}-\frac{A_{1}^{2}}{A_{0}}\right).
\end{equation}
Within the Kubo formalism \cite{Mahan} the kinetic coefficients
are given in terms of equilibrium state current-current
correlation functions of the particle and the heat currents in
the system.  Namely we have
\begin{equation}
  \label{eq:A_coefficients}
  A_{m} = \beta^{m}\lim_{\omega\rightarrow 0}Z_{m}(i\nu \rightarrow
  \omega+i0)\ ,
\end{equation}
where
\begin{eqnarray}
  \label{eq:Z0_function}
  Z_{0}(i\nu) &=& \frac{i\hbar}{i\nu\beta}\int_{0}^{\beta}
  d\tau e^{i\nu\tau}\langle T_{\tau}j^{x}(\tau)j^{x}(0)\rangle \ ,
  \\
  \label{eq:Z1_function}
  Z_{1}(i\nu) &=& \frac{i\hbar}{i\nu\beta}\int_{0}^{\beta}
  d\tau e^{i\nu\tau}\langle T_{\tau}j^{x}(\tau)Q^{x}(0)\rangle \ ,
  \\
  \label{eq:Z2_function}
  Z_{2}(i\nu) &=& \frac{i\hbar}{i\nu\beta}\int_{0}^{\beta}
  d\tau e^{i\nu\tau}\langle T_{\tau}Q^{x}(\tau)Q^{x}(0)\rangle \ .
\end{eqnarray}

To evaluate these correlation functions, an expression for
electric and heat currents, $j^{x}$ and $Q^{x}$, are needed. Once
those currents are evaluated, then calculation of the transport
properties within  DMFT reduces to the evaluation of the transport
function
\begin{equation}
  \label{eq:transportfunction}
  \phi^{xx}(\epsilon) = \frac{1}{\cellvol}\sum_{k}
  \trace
  \left\{
    {v}^{x}_{k}(\epsilon){\rho}_{k}(\epsilon)
    {v}^{x}_{k}(\epsilon){\rho}_{k}(\epsilon)
  \right\},
\end{equation}
and the transport coefficients
\begin{eqnarray}
  \label{eq:A_coefficients2}
  A_{m} =  N_{spin}\pi\hbar
  \int_{-\infty}^{\infty}d\epsilon\phi^{xx}(\epsilon)
  f(\epsilon)f(-\epsilon)(\beta\epsilon)^{m},
\end{eqnarray}

The momentum integral in Eq.~(\ref{eq:transportfunction}) extends
over the Brillouin zone, $\cellvol$ is the volume of the unit
cell. The simplest form of the velocity is $\la
k\beta|\frac{1}{m}\nabla_x|k\alpha\ra =v_{k}^{\alpha\beta}$ and
it requires evaluation of matrix elements of $\nabla_x$. However
an alternative form of the current and the transport function can
be derived via the Peirls substitution generally in the
non-orthogonal basis and is described in
Appendix~\ref{app:transport}. These two procedures generally give
different answers ~\cite{Blount:1962,Paul:2002,Millis:2001}.

Next we define the energy dependent velocity as
\begin{equation}
  \label{eq:totalvelocity}
  \vec{{v}}_{k}(\epsilon) = \vec{{v}}_{k} - \epsilon\vec{{u}}_{k}.
\end{equation}
The second term is due to the non-orthogonality of the basis or
more specifically due to overlap between orbitals at different
sites, local non-orthogonality does not contribute to the
velocity.  The spectral density matrix ${\rho}_{k}(\epsilon)$ is
the multiorbital generalization of the regular single orbital
density of states and is given in terms of the retarded Green's
function, ${G}$, of the system by the equation
\begin{equation}
  \label{eq:spectraldensity}
  {\rho}_{k}(\epsilon) = -\frac{1}{2\pi}
  \left\{
    {G}_{k}(\epsilon)-[{G}_{k}(\epsilon)]^{\dagger}
  \right\}.
\end{equation}
Finally the Green's function is given by
\begin{equation}
  \label{eq:Greensfunction}
  {G}_{k}(z) =
  \left[
    (z+\mu){O}_{k}-{H}^{0}_{k}-{\Sigma}(z)
  \right]^{-1}.
\end{equation}

Note here, that in accordance with the DMFT the self-energy
matrix is assumed momentum independent.  Now given an effective
Hamiltonian for the system, an overlap matrix and the self-energy,
the equations above give a complete prescription for computing
the transport parameters. For computation of
Eq.~(\ref{eq:transportfunction}) we have developed two methods,
one method  generalizes the analytical tetrahedron method (ATM)
\cite{Lambin:1984} and the other one uses one-particle GF method
in DMFT ~\cite{Anisimov:1997}, used to compute spectral densities
in band structure calculations. First the total Hamiltonian,
${H}_{k}(\epsilon) = {H}_{k}^{0}+{\Sigma}(\epsilon)$ is
diagonalized and written in the form
\begin{equation}
  \label{eq:diagonalrepresentation}
  {H}_{k}(\epsilon) =
  {O}_{k}
  {A}^{R}_{k}(\epsilon)
  {E}_{k}(\epsilon)
  {A}^{L}_{k}(\epsilon)
  {O}_{k},
\end{equation}
where ${E}_{k}$ is the diagonal matrix of complex eigenvalues and
${A}^{R}_{k}$ and ${A}^{L}_{k}$ are the right and the left
eigenvector matrices respectively.  Then the Green's function can
be written as
\begin{equation}
  \label{eq:diagonalGreeen}
  {G}_{k}(\epsilon) =
  {A}^{R}_{k}(\epsilon)
  [(\epsilon+\mu){I}-{E}_{k}(\epsilon)]^{-1}
  {A}^{L}_{k}(\epsilon),
\end{equation}
with ${I}$ being the identity matrix. The transport function can
now be expressed as
\begin{eqnarray}
  \label{eq:rst_representation}
  \phi^{xx}(\epsilon) &=&
  -\frac{1}{2\pi^{2}\cellvol}\real\sum_{k,pq}
  \Bigl\{
    r^{x}_{k,qp}r^{x}_{k,pq}
    D_{k,p}D_{k,q}
  \Bigr.
  \nonumber \\
  \Bigl.
   & -&\frac{1}{2}
    \left[
      s^{x}_{k,qp}t^{x}_{k,pq}+
      s^{x}_{k,pq}t^{x}_{k,qp}
    \right]
    D_{k,p}(D_{k,q})^{*}
  \Bigr\},
\end{eqnarray}
where the matrices ${r}^{x}$, ${s}^{x}$ and ${t}^{x}$ are
\begin{eqnarray}
  \label{eq:rst_definition}
  {r}^{x}_{k} &=& {r}^{x}_{k}(\epsilon) \equiv
  {A}^{L}_{k}(\epsilon)
  {v}^{x}_{k}(\epsilon)
  {A}^{R}_{k}(\epsilon),
  \\\nonumber
  {s}^{x}_{k} &=& {s}^{x}_{k}(\epsilon) \equiv
  {A}^{L}_{k}(\epsilon)
  {v}^{x}_{k}(\epsilon)
  [{A}^{L}_{k}(\epsilon)]^{\dagger},
  \\\nonumber
  {t}^{x}_{k} &=& {t}^{x}_{k}(\epsilon) \equiv
  [{A}^{R}_{k}(\epsilon)]^{\dagger}
  {v}^{x}_{k}(\epsilon)
  {A}^{R}_{k}(\epsilon),
\end{eqnarray}
and ${D}_{k}$ is a diagonal matrix defined by
\begin{equation}
  \label{eq:denominatormatrix}
  {D}_{k} = {D}_{k}(\epsilon) \equiv
    [(\epsilon+\mu){I}-{E}_{k}(\epsilon)]^{-1}.
\end{equation}
When the computation of the transport function is carried out one
is faced with computing integrals of the form
\begin{eqnarray}
  \label{eq:tetrahedronintegrals}
  &&
  \sum_{k}\frac{r^{x}_{k,pq}r^{x}_{k,qp}}
  {(\epsilon+\mu-E_{k,p})(\epsilon+\mu-E_{k,q})},
  \\\nonumber &&
  \sum_{k}\frac{s^{x}_{k,pq}t^{x}_{k,qp}}
  {(\epsilon+\mu-E_{k,p})(\epsilon+\mu-E_{k,q}^{*})}.
\end{eqnarray}
\noindent The strategy that is used to compute these integrals is
similar in spirit to the analytical tetrahedron method.  The
Brillouin zone is split up into a collection of equal sized
tetrahedra and the integral over each tetrahedron is taken using
linear interpolation between the four corners of the
tetrahedron.  In the analytical tetrahedron method the numerator
and the energy eigenvalues in the denominator are linearized
independently and the resulting integral is then done
analytically.  In our case we would want to follow the same rule
which results in two linear functions in the denominator.
Unfortunately we have not been able to evaluate that integral in
the most general case, i.e.~when none of the tetrahedron corners
are degenerate although solutions can be found for degenerate
cases when at least two of the four corners of the tetrahedron are
identical.  Hence we have to pursue further approximations which
we outline below.

The two main integrals that we need to compute are of the form
\begin{eqnarray}
  \label{eq:tetrahedronintegrals2}
  T_{SS}^{pq} &=&
  \sum_{k\in\Delta}\frac{F(k)}{(z-E_{k,p})(z-E_{k,q})},
  \\\nonumber
  T_{OS}^{pq} &=&
  \sum_{k\in\Delta}\frac{F(k)}
  {(z-E_{k,p})(z-E_{k,q})^{*}}.
\end{eqnarray}

\noindent Here $\Delta$ denotes the tetrahedron and $SS$
indicates that the imaginary parts of both denominators have the
same sign and $OS$ indicates that they have the opposite sign.
This is ensured by the fact that the self-energy is retarded and
$z$ is real.  For the diagonal case $(p = q)$ the $T_{SS}$
integral can be computed exactly by linearizing the eigenvalues
in the denominator, one simply needs to differentiate the ATM
formulas by Lambin and Vigneron \cite{Lambin:1984}. For the
diagonal $T_{OS}$ however we note that the numerator is real and
therefore we can  write the integral in the following form
\begin{equation}
  T_{OS}^{pp} =
  \imag\sum_{k\in\Delta}\left(\frac{F(k)}{\gamma_{k,p}}\right)
  \frac{1}{z-E_{k,p}},
\end{equation}
where $\gamma_{k,p} = \imag E_{k,p}$.  We note that
$\gamma_{k,p}$ is solely due to the self-energy, which is
momentum independent and thus it is reasonable to expect that
$\gamma_{k,p}$ changes little with momentum.  Hence the term in
the parenthesis will be approximated linearly within the
tetrahedron and the resulting integral can be computed with the
ATM.

The off-diagonal case $(p \neq q)$ for both $T_{SS}$ and $T_{OS}$
is treated the same way so we will just look at $T_{SS}$.  Both
factors in the denominator are inspected and we determine which
one has larger modulus (on average if necessary).  Then we write
the integral on the form
\begin{equation}
  \left.T_{SS}^{pq}\right|_{p\neq q} =
  \sum_{k\in\Delta}\left(\frac{F(k)}{(z-E_{k})_{L}}\right)
  \frac{1}{(z-E_{k})_{S}},
\end{equation}
where $L$ indicates the denominator with the larger modulus and $S$ indicates
the one with the smaller modulus.  The term in the parenthesis is now
approximated linearly within the tetrahedron and the resulting integral can be
computed with the ATM.

The approach described here to compute the transport function has
been tested numerically against models where other methods can be
used to evaluate the transport function.  For cubic systems with
nearest neighbor hopping one can for instance evaluate both the
density of states and the transport function quite efficiently
using Fast Fourier Transforms \cite{Georges:1996}.  In general
the results are quite accurate.

\subsection{Small scattering limit}%
\label{sec:small_scattering}

In order to make connections with previous approaches to the
computation of transport properties it is interesting to consider
the small scattering limit.  So we take the self-energy of the
form
\begin{equation}
{\Sigma}(\epsilon) =
{\Sigma}'(\epsilon)+\gamma{\Sigma}''(\epsilon),
\end{equation}
where ${\Sigma}'(\epsilon)$ is the real part of the self-energy
matrix, $\gamma{\Sigma}''(\epsilon)$ is the imaginary part and
$\gamma$ is a small parameter.

It is clear that the transport function will diverge as $1/\gamma$
and thus we can approximate the numerator matrix elements to
zeroth order in $\gamma$.  Within this approximation the
transport function can be written as
\begin{equation}
\label{eq:smallscatteringlimit}
\phi^{xx}(\epsilon) =
\frac{1}{\cellvol}\sum_{k,p}
(v^{x}_{k,p})^{2}\tau_{k,p}(\epsilon)\delta(\epsilon+\mu-E_{k,p}'),
\end{equation}
where $E_{k,p}'$ are the eigenvalues of
${H}^{0}_{k}+{\Sigma}'(\epsilon)$ and $v^{x}_{k,p}$ denotes the
corresponding band velocity.  The lifetime $\tau_{k,p}(\epsilon)$
is formally given by
\begin{equation}
\tau_{k,p}(\epsilon) = \frac{1}{2\pi|\imag E_{k,p}|},
\end{equation}
here $E_{k,p}$ are the eigenvalues of the full Hamiltonian. The
imaginary part of these eigenvalues is due to the scattering term
and is therefore to first approximation linear in $\gamma$. The
lifetime therefore diverges as $1/\gamma$ but for a finite value
of $\gamma$ we regard Eq. (\ref{eq:smallscatteringlimit}) as an
approximation to the transport function and we will refer to this
approach as the small scattering approximation.

In spite of the limited validity of the small scattering
approximation it is useful in the sense that it is
computationally much simpler to evaluate the transport function
in the small scattering approximation than in the general case.
Therefore it can be used in order to obtain a rough idea of the
behavior of the transport parameters.

The equations of the small scattering approximation are very
similar to the formulae that have been used by other groups to
compute the transport parameters of real materials
\cite{Singh:1997,Kim:1997,Fornari:1999}.  In particular the
assumption of constant lifetime is quite often used in practice,
especially when the thermopower is being calculated.  In this
case we obtain
\begin{equation}
\label{eq:constantscatteringlimit}%
 \phi^{xx}(\epsilon) = \tau\Phi^{xx}(\epsilon),
\end{equation}
where the so called transport density, $\Phi$, is defined as
\begin{equation}
\Phi^{xx}(\epsilon) =
\frac{1}{\cellvol}\sum_{k,p}
(v^{x}_{k,p})^{2}\delta(\epsilon+\mu-E_{k,p}').
\end{equation}

Numerical tests have shown that while the small scattering
approximation can be quite good for broad bands it does not work
well in narrow bands such as the dynamically generated
quasiparticle bands of strongly correlated systems due to
constant time approximation used.

In the case of the thermopower we obtain
\begin{eqnarray}
S &=& -\frac{\kb}{|e|} \left(\frac {\int_{-\infty}^{\infty}
\Phi^{xx}(\epsilon)f(\epsilon)f(-\epsilon)(\beta\epsilon)}
{\int_{-\infty}^{\infty}
\Phi^{xx}(\epsilon)f(\epsilon)f(-\epsilon)} \right)
\nonumber \\
&\stackrel{T \longrightarrow 0}{=}&
-\frac{\kb}{|e|}\frac{\pi^{2}\kb T}{3} \left.
\frac{d}{d\epsilon}\ln\Phi^{xx}(\epsilon) \right|_{0}.
 \label{eq:S_Mott}
\end{eqnarray}

This of course is the classical Mott relation for the
thermopower. In the literature this equation is often quoted with
the transport density replaced by the spectral density and much
emphasis placed on the fact that in case the Fermi-level
coincides with a Van Hove singularity the thermopower diverges.
This conclusion is not supported when the correct form for the
thermopower is used since no Van Hove singularities are present
in the transport density.

For free electrons the transport density is given by
\begin{equation}
  \label{eq:freeElectronPhi}
  \Phi^{xx}(\epsilon) =
  \frac{1}{12\pi^{2}}
  \left(\frac{2m_{e}}{\hbar^{2}}\right)^{3/2}
  \epsilon^{3/2},
\end{equation}
and therefore we get
\begin{equation}
  \label{eq:freeElectronS}
  S = -\frac{\kb}{|e|}\frac{\pi^{2}\kb T}{2}
  \frac{1}{\epsilon_{F}} = - n^{-2/3}T \times 0.281\frac{nV}{K},
\end{equation}
where the density, $n$, is measured in electrons per cubic Bohr
radius and the temperature, $T$, is measured in Kelvin. In case
the effective mass of the electrons is enhanced the thermopower
will simply increase by the enhancement factor.

The enhancement of the thermopower can also be deduced from the
Mott equation in case the only effect of the real part of the
self-energy is to change the effective mass of the bands that
cross the Fermi-surface.  If we assume that the change in
effective mass is the same for all the bands that participate in
the transport the low-temperature thermopower becomes
\begin{eqnarray}
S \simeq -\frac{\kb}{|e|}\frac{\pi^{2}\kb T}{3Z} \left.
\frac{d}{d\epsilon}\ln\Phi^{0,xx}(\epsilon) \right|_{0},
\end{eqnarray}
where the non-interacting transport density
$\Phi^{0,xx}(\epsilon)$ is defined by
\begin{equation}
\Phi^{0,xx}(\epsilon) =
\frac{1}{\cellvol}\sum_{k,p}
(v^{0.x}_{k,p})^{2}\delta(\epsilon+\mu-E^{0}_{k,p}).
\end{equation}
Here $Z$ denotes the quasiparticle residue of the bands
involved.  Hence we see indeed that the low-temperature
thermopower is enhanced by a factor of $1/Z$ compared to the
non-interacting thermopower.

To summarize theoretical part of the paper we provide the
computational scheme used to obtain the transport properties
\begin{equation*}
LDA\stackrel{H_{LDA}}{\longrightarrow }
DMFT\stackrel{H_{LDA}+\Sigma(\omega)}{\longrightarrow }
TRANSPORT,%
\end{equation*}
i.e.
\begin{itemize}
\item  run LDA program  to obtain the Kohn-Sham (LDA) Hamiltonian
\item  make tight-binding calculations (or downfolding) to extract kinetic
       part of the effective Hamiltonian
\item  construct the Hubbard-like effective Hamiltonian adding the Coulomb
       repulsion to the kinetic part of the effective Hamiltonian
\item  run DMFT-QMC(SUNCA) solver to obtain the self-energy  for the effective Hamiltonian
\item  construct  LDA+DMFT Hamiltonian upfolding the self-energy
       to the LDA Hamiltonian
\item  run transport program with the LDA+DMFT Hamiltonian.
\end{itemize}

\section{Test System and DMFT Results}%
\label{sec:dmft_results}

\subsection{Test System}

To test obtained transport equations on a realistic system we
decided to use doped LaTiO$_3$ compound. The
La$_{1-x}$Sr$_{x}$TiO$_{3}$ series has been studied very
extensively in the past \cite{Sunstrom:1992, Maeno:1990,
Crandles:1992, Onoda:1997a, Onoda:1998, Hays:1999} and can be
regarded as being one of the prime examples exhibiting the
Mott-Hubbard metal-insulator transition.  The end compound
LaTiO$_{3}$ when prepared well is a Mott-Hubbard insulator
although in the literature it is often characterized as a
correlated or a poor metal. At high temperature this material is
paramagnetic. The other end compound SrTiO$_{3}$ is an
uncorrelated band insulator with a direct gap of 3.3~eV.
Electronic structure properties of the
La$_{1-x}$Sr$_{x}$TiO$_{3}$ series is governed by the triple
degenerate cubic $t_{2g}$ bands of the $3d$ orbitals ($d^1$ ionic
configuration)~\cite{Pavarini:2003:CM}. In the distorted
structure of LaTiO$_{3}$ the degeneracy of the band has been
lifted and the single electron occupies a very narrow,
non-degenerate $d_{xy}$ band~\cite{Goodenough:1971}. Studies of
the magnetic susceptibility do indeed indicate that the
electronic structure of the $Pbnm$ phase is that of a narrow
$d_{xy}$ band, which then with doping changes into a broad
$t_{2g}$ band (calculated bandwidth is $W=2.7$~eV) with
degenerate $d_{xy}$, $d_{xz}$ and $d_{yz}$ orbitals in the $Ibmm$
and $Pm3m$ phases. As a function of doping the material behaves
as a canonical doped Mott insulator. The specific heat and the
susceptibility are enhanced, the Hall number is unrenormalized
while the photoemission spectral function has a resonance with a
weight that decreases as one approaches half filling. Very near
half filling, (for dopings less than 8~\%) the physics is fairly
complicated. At small doping  an antiferromagnetic metallic phase
is observed ~\cite{Kumagai:1993, Okada:1993, Onoda:1998}.

To obtain LDA band structure of LaTiO$_3$ we used the liner
muffin-tin orbitals (LMTO) method in its atomic sphere
approximation (ASA) with the basis Ti$(4s,4p,3d)$, O$(2s,2p)$
and  La$(6s,5p,5d)$ assuming for simplicity instead a real
orthorhombic structure with a small distortions a cubic one with
the same volume and the lattice constant $a_0=7.40$~a.u.. This
approximation brings to a slight overestimation of the effective
bandwidth and underestimation of the band gap between valence and
conduction bands. In photoemission studies of LaTiO$_3$
\cite{Nekrasov:2000} similar basis has been used.

Using LDA band structure one can compute and compare with
experiment the linear coefficient of specific heat which is
simply given in terms of the density of states at the Fermi level
by
\begin{equation}
\label{eq:gamma} \gamma = 2.357\left[\frac{mJ}{mol K^2}\right]
\frac{\rho_{tot}(E_f)[\mbox{states/(eV unitcell)}]}{Z},
\end{equation}
where $Z$ is the quasiparticle residue or the inverse of the mass
renormalization. In LDA calculations the value of $Z$ is equal to
one. Doping dependence of the linear coefficient of specific heat
in LDA calculations was computed within the rigid band model. Our
results along with the experimental data are presented in
Table~\ref{tb:LDA_gammas}.
\begin{table}[h]
\begin{center}
\begin{tabular}{|c|c|c|}     \hline
Doping &Experiment& LDA   \\ \hline \hline
  5\%  &  16.52   & 3.23  \\ \hline
 10\%  &  11.51   & 3.16  \\ \hline
 20\%  &   8.57   & 3.00  \\ \hline
 30\%  &   7.70   & 2.82  \\ \hline
 40\%  &   6.21   & 2.67  \\ \hline
 50\%  &   5.38   & 2.52  \\ \hline
 60\%  &   4.55   & 2.38  \\ \hline
 70\%  &   4.35   & 2.19  \\ \hline
 80\%  &   3.52   & 2.10  \\ \hline
\end{tabular}
\end{center}
\caption{The linear coefficient of specific heat, $\gamma$, for
La$_{1-x}$Sr$_{x}$TiO$_{3}$  measured in units of $\frac{mJ}{mol
K^{2}}$.  The experimental data is taken from
\protect\cite{Tokura:1993:PRL}. LDA  data for the linear
coefficient of specific heat  are computed from LaTiO$_{3}$ LDA
DOS.} \label{tb:LDA_gammas}
\end{table}

In general, we see that the LDA data for $\gamma$ are lower than
the experimental values, indicating a strong mass renormalization.
We note also that as we get closer to the Mott-Hubbard transition
the effective mass grows significantly.  What is consistent with
DMFT modeling of the Mott-Hubbard transition which shows that
indeed the effective mass diverges at the transition. We should
note however that this is not a necessary signature for the
Mott-Hubbard transition: in V$_{2}$O$_{3}$ the pressure driven
metal-insulator transition is accompanied by the divergence of
the effective mass whereas the doping driven transition in the
same system does not show that divergence \cite{Carter:1993}.

The physical picture of the studied material is quite transparent,
very near half filling (dopings less than 8~\%) the Fermi energy
becomes very small and now is comparable with the exchange
interactions and structural distortion energies. A treatment
beyond single site DMFT then becomes important to treat the spin
degrees of freedom. On the other hand for moderate and large
doping, the Kondo energy is the dominant energy and DMFT is
expected to be accurate. This was substantiated by a series of
papers which compared DMFT calculations in a single band or
multiband Hubbard models using simplified density of states with
the physical properties of real materials. Ref.
~\onlinecite{Rozenberg:1994} addressed the enhancement of the
magnetic susceptibility and the specific heat as half filling is
approached. The optical conductivity and the suppression  of the
charge degrees of freedom as the Mott insulator is approached was
described in Refs.~\onlinecite{Rozenberg:1995,Kotliar:1996}, the
observation that the Hall coefficient is not renormalized was
found in Refs.~\onlinecite{Kajueter:1996a,Kajueter:1997}. Finally
the thermoelectric power on model level using iterative
perturbation theory (IPT) as impurity solver was investigated by
P\'alsson et al.~\cite{Palsson:1998}.

Given the fact that only very simple tight-binding
parameterizations were used in those works, and the fact that a
large number of experiments were fit with the same value of
parameters one should regard the qualitative agreement with
experiment as very satisfactory. The photoemission spectroscopy
of this compound, as well as in other transition metal compounds
are not completely consistent with the bulk data, and it has been
argued that disorder, and modeling of the specific surface
environment is required to improve the agreement with experiment
~\cite{Sarma:1996}. In this situation, it is clear that this is
the simplest system for study, and it was in fact the first
system studied by LDA+DMFT ~\cite{Anisimov:1997}.

The important  questions to be addressed are the degree of
quantitative accuracy of DMFT. Furthermore, given the simplicity
of this system, and the existence of well controlled experiments,
it is an ideal system for testing the effects of
different approximations within the LDA+DMFT scheme.%

\subsection{The model} %
\label{sec:model}

As we pointed out in Sec.~\ref{sec:dmft} for a correct description
of a system with strong electron correlations one needs to bring
the self-energy into the heavy orbitals. For this purpose a model
which correctly describes the physics of interacting orbitals is
needed. In this paper we consider a three-band Hubbard model
which underlying non-interacting dispersion relation is that of
the degenerate cubic $t_{2g}$ band of the transition metal $3d$
orbitals. For simplicity the Hubbard interaction term is taken to
be $SU(6)$ invariant i.e. there is equal interaction between two
electrons of opposite spin in the same orbital as there is
between two electrons in different orbitals on the same site. The
more general case will be revisited in future publications.

Value of the interaction strength in our model is chosen large
enough to exhibit  metal-insulator behavior in the studied
compound. In units of half bandwidth, $D$, the interaction
strength is taken to be equal $U=5D$. The interaction strength
should be regarded as an input parameter which value has be
adjusted to the experimental situation. Saying this, we mean the
chosen interaction strengths should be good enough to reproduce
as many physical properties as only possible with maximum
proximity to experiment. To investigate dependence of calculated
physical properties on the interaction strength we calculate all
properties studied in this paper for two values of Coulomb
repulsion $U=3D$ and $5D$. On the model level $U=4D$ is the value
very close to the minimum interaction to get metal-insulator
transition (MIT) for integer filling $n=1$ even in three-fold
degenerate Hubbard model using DMFT as an instrument which takes
care of the interaction in the system. Hence our choice of the
interaction should guarantee exploration of two physically
different  behaviours of the system with and without the MIT. In
literature absolute value of Coulomb interaction is magnitude
under discussion mainly because there is no direct and reliable
method to extract it neither experimentally no theoretically.
The  uncertainty between different theoretical
methods~\cite{Solovyev:1996,Nekrasov:2000} attempted to estimate
value of $U$ is quite substantial ranging the interaction
strength from 3.2~eV to 5~eV.

It should be noticed that although this choice of parameters is
consistent with insulating behavior of this system it might have
limited validity in the real system at low doping. Since
La$_{1-x}$Sr$_{x}$TiO$_{3}$ is known to undergo several
structural transformations upon doping and in particular the
structure of LaTiO$_{3}$ is distorted away from the cubic
perovskite structure and in fact the distortion lifts the
degeneracy of the t$_{2g}$ orbitals and the groundstate orbital
is a narrow non-degenerate $d_{xy}$ orbital.  Hence one might
expect that the Mott-Hubbard transition in this system would be
better described in a one-band model $(x<0.08)$.  At larger
dopings $(x>0.08)$ it is however clear that the system is
degenerate and thus our model can be expected to give a reasonably
good description in the larger doping range. In the present paper
we do not consider the effect of lifting the degeneracy due to
Jahn-Teller distortion rather we explore the tree-fold degenerate
Hubbard model in the whole region of doping interval including
$n=1$ point.

\begin{figure}
\centering
\centering \epsfig{file=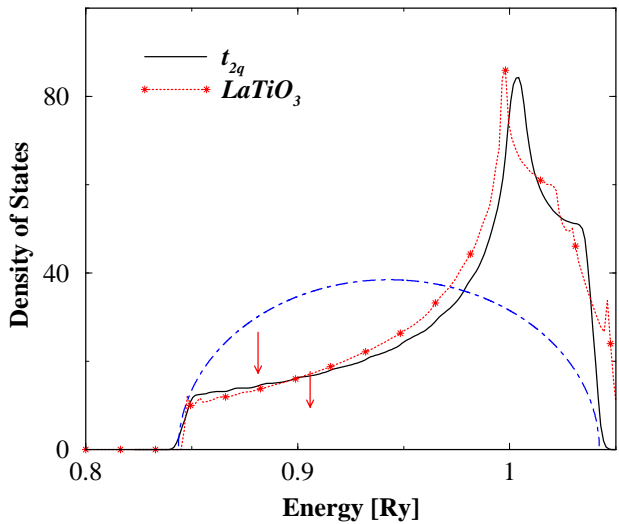,angle=0,width=3.2in}\\
\caption{LDA DOS of LaTiO$_3$ (dotted line with star symbols),
its tight-binding fit(solid line) and semicircular DOS
(dot-dashed line). Arrows indicate Fermi level position for
filling $n=0.8$ (the first one is for the semicircular DOS, and
the second one is for the tight-binding fit).
}%
\label{fig:realdoses}
\end{figure}

The kinetic part of the model Hamiltonian has been obtained from
tight-binding LMTO ASA calculations. The band structure of the
compound around the Fermi level consists of three-fold degenerate
Ti~$3d$~$t_{2g}$ band, hosting one-electron, which is a rather
well separated from an empty Ti $3d$ $e_g$ band located above
$t_{2g}$ band. A rather broad gap below $t_{2g}$ separates Ti
$3d$ and completely filled $2p$ oxygen band. Hence it is quite
straight forward to make the tight-binding fit of $t_{2g}$ band
to be used in in the impurity solvers. To achieve asymmetry in
tight-binding DOS one needs take into account the next nearest
neighbors, so called, $t'$ term on Ti sublattice. The dispersion
which we obtained from the fit is the following: $\epsilon_{\bf
k}=2 t (\cos k_x + \cos k_y) +2 t' \cos(k_x+k_y) + 2 t_\perp \cos
k_z$, where $t=-0.02424$, $t'=-0.006$, $t_{\perp}=-0.00151$ in Ry
units. $t_{2g}$ part of LaTiO$_3$ DOS (dotted line) and its fit
(solid line) are presented in Fig.~\ref{fig:realdoses}. We also
added one more curve in Fig.~\ref{fig:realdoses} corresponding to
semicircular DOS which we will use for different kind of
benchmarking of our approach.

Making tight-binding fit of Ti~$3d~t_{2g}$ bands in LaTiO$_3$ we
effectively do downfolding of the whole Hamiltonian (better to say
the main part of the Hamiltonian without Ti~$3d~t_{2g}$ bands) of
the compound onto three $t_{2g}$ bands i.e. we incorporate
information of the whole Hamiltonian into part of it. In  this
case $H_{eff} $ and $O_{eff}$ are $3\times 3$ matrices. $H_{hh}$
and $O_{hh}$ are parts of the Hamiltonian and the overlap matrix
corresponding to $t_{2g}$ block of the original Hamiltonian.
$H_{ll}$ and $O_{ll}$ are parts of the Hamiltonian and the
overlap matrices of the Hamiltonian without $t_{2g}$ block.

\begin{figure}[h]
\centering
\epsfig{file=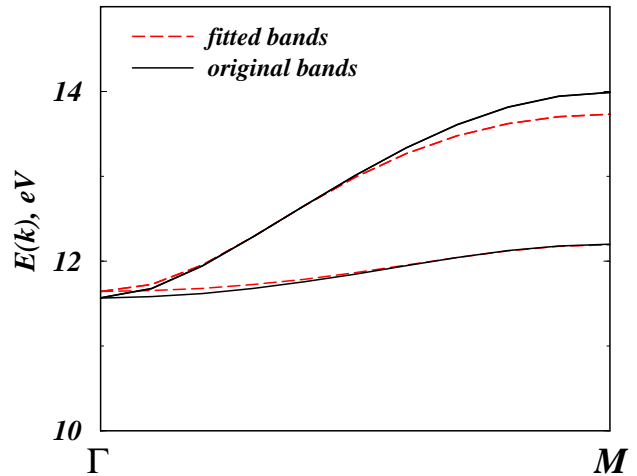,angle=0,width=3.2in} %
\caption{t$_{2g}$ bands of real band structure of LaTiO$_3$ (solid
lines) and downfolded bands obtained from Eq.~(\ref{eq:h_eff}).
The Fermi level is at $\varepsilon_F = 12.54$~eV.
}%
\label{fig:h_eff}
\end{figure}

Our results of downfolding are presented in Fig.~\ref{fig:h_eff}.
We plotted bands of the whole Hamiltonian which correspond to
Ti~$3d~t_{2g}$ bands by solid lines (the upper line is double
degenerate one) and bands obtained from Eq.~(\ref{eq:h_eff}) by
dashed lines.
We see that bandwidth of $t_{2g}$ bands obtained from
Eqs.~(\ref{eq:h_eff}) is only slightly smaller than the original
one. The downfolding reproduces original band structure
one-to-one in approximately 1~eV energy window around the
chemical potential. If one uses only $H_{hh}$ and $O_{hh}$
matrices (not effective $H_{eff}$ and $O_{eff}$ but simply parts
of the total Hamiltonian corresponding to $t_{2g}$ block)  one
gets nearly dispersionless levels only as it is naturally
expected.

\begin{figure}[h]
\centering \epsfig{file=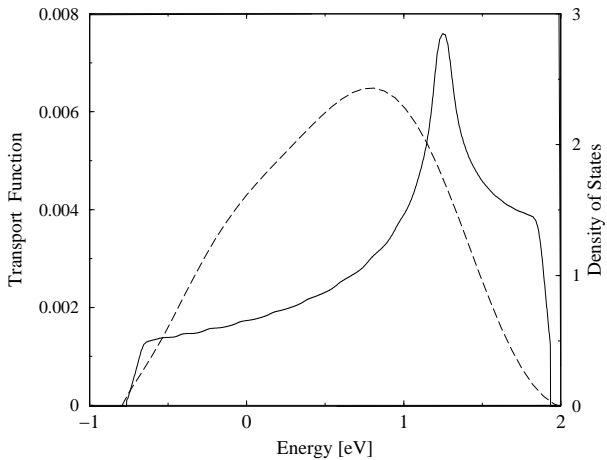,angle=-90,width=3.2in}
\caption{The transport function for the LDA tight-binding
Hamiltonian computed within the small scattering approximation
with a constant lifetime for all orbitals along with DOS for
$n=0.8$. Zero energy corresponds to the zero temperature
Fermi-level.}%
\label{fig:LDA_PHI}
\end{figure}

As the last step in this section we calculate the transport
function and the thermoelectrical power for three-fold degenerate
tight-binding model. The transport function that we obtain for
this model is shown in Fig.~\ref{fig:LDA_PHI}.  We note that the
slope of the transport function in the hole doping $(\mu <0.0)$
regime is positive and therefore the LDA evaluation of the
thermopower results in electron-like thermopower.  The results
for a few chosen doping values computed at temperature equal to
300~K are displayed in table \ref{table:LDA_S} where we have also
displayed the experimental results from
Ref.~\onlinecite{Hays:1999}.

\begin{table}[h]
\begin{center}
\begin{tabular}{|c|c|c|}\hline
Doping   &  Experiment   &  LDA data  \\ \hline
  5\%     &     -5.2      &    -5.6   \\ \hline
 25\%     &     -9.3      &    -7.8   \\ \hline
 50\%     &    -18.3      &    -9.3   \\ \hline
 75\%     &    -29.4      &   -18.2   \\ \hline
 80\%     &    -41.2      &   -22.8    \\\hline
\end{tabular}
\end{center}
\caption{The thermopower, $S$, of La$_{1-x}$Sr$_{x}$TiO$_{3}$ at
300K measured in units of ${\mu V}/{K}$ is computed using LDA
band structure. The experimental data are taken
from~\protect\cite{Hays:1999}.} \label{table:LDA_S}
\end{table}

It is quite noteworthy that for the two lowest doping values in
the table the LDA and the experiment are in a good agreement. For
higher values of doping however the experimental values are about
twice as large as the LDA values. The good agreement at low
doping should be regarded as mostly accidental one since the
experimental data for doping values less than 5\% show the
hole-like thermopower which the LDA of course will not be able to
reproduce.

\subsection{Summary of DMFT results}%
\label{sec:dmft_qmc_res}

In the previous section we described how to obtain the Hubbard
like Hamiltonian with kinetic part coming from downfolded bands
and interaction part defined by renormalized Coulomb repulsion. In
this section we address the issue how to compute and what is the
main effect of the second part of the Hamiltonian, namely
interactions, on physical properties of the studied system. The
method which we used to solve the Hamiltonian is the dynamical
mean-field theory which we described in section~\ref{sec:dmft}.

\begin{figure}
\centering
\centering \epsfig{file=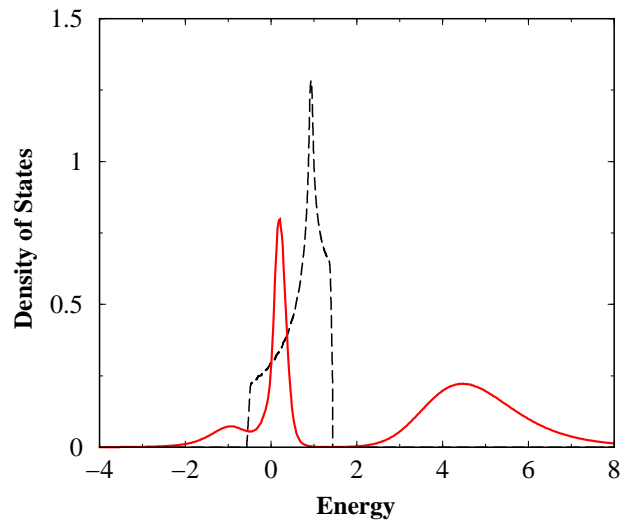,angle=0,width=3.2in}\\
\caption{Tight-binding (dashed lines) and renormalized (solid
line) DOSes for filling n=0.8 and $\beta=32$ at $U=5D$.
}%
\label{fig:doses}
\end{figure}

Strong correlations dramatically renormalize tight-banding LDA
DOS which we plot in Fig.~\ref{fig:doses} by dashed line. As it
is seen from the plot it has bandwidth $2D$ and very asymmetric
shape. In the same figure it is shown by solid line the
renormalized DOS obtained from DMFT-QMC calculations at
temperature $T=980$~K. Instead one QP peak of the non-interacting
$t_{2g}$ DOS we have nearly classical Hubbard DOS picture where
the central peak of the non-renormalized DOS is redistributed
between two Hubbard subbands and strongly renormalized central
peak. Although the total bandwidth increases substantially, width
of the central peak gets strongly reduced. We also notice the
steepness of renormalized DOS at the chemical potential
indicating reduction of the quasiparticle residue which is in LDA
equal to one.

\begin{figure}
\centering
\epsfig{file=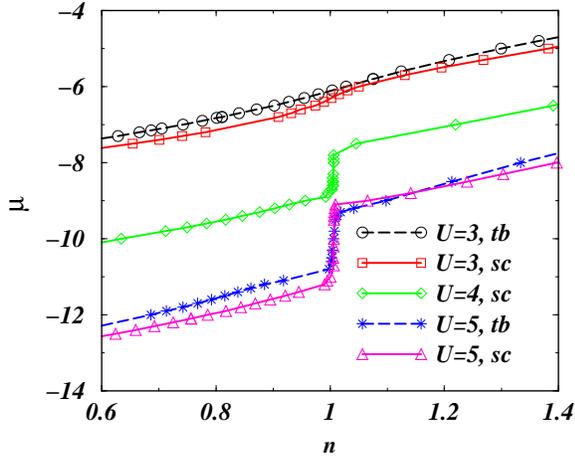,angle=0,width=3.0in}\\
\caption{The chemical potential, $\mu$, versus  filling, $n$ for
semicircular (sc) and tight-binding (tb) DOSes and various values
of interaction, $U$, and temperature $\beta = 16$.
}%
\label{fig:mu_n}
\end{figure}

\begin{figure}
\centering
\epsfig{file=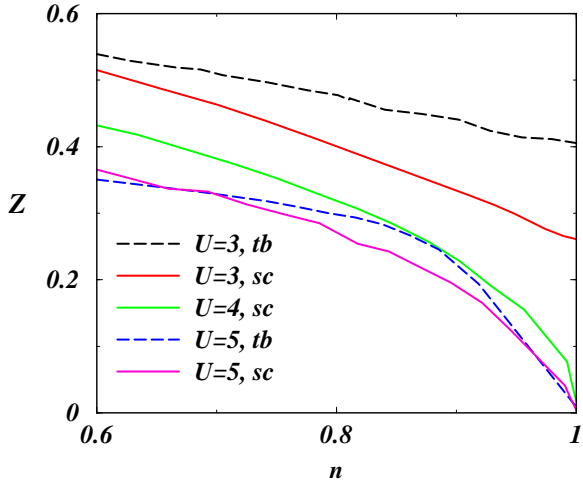,angle=0,width=3.0in}\\
\caption{Density dependence of the quasiparticle residue, $Z$,
for semicircular (sc) and tight-binding (tb) DOSes and various
values of interaction, $U$ and temperature $\beta = 16$.
}%
\label{fig:z_n}
\end{figure}

So, the main effect expected from electron interactions is to
reproduce the Mott transition when the system approaches an
integer filling. One can see indications of the MIT in filling,
$n$, dependencies of the chemical potential, $\mu$, and
quasiparticle residue, $Z$. The MIT is clearly indicated by jump
of $\mu$ versus $n$ dependence (the chemical potential changes
while filling remains the same) plotted in Fig.~\ref{fig:mu_n}
and also by vanishing energy scale seen in $Z$ versus $n$
dependence in Fig.~\ref{fig:z_n} while approaching the Mott
transition.

In Fig.~\ref{fig:mu_n} we plot the chemical potential against
filling around filing $n=1$ for three values of Coulomb
interaction $U=3D,4D$ and $5D$ in units of the half bandwidth and
for two shapes of DOS (semicircular and tight-binding). We notice
here that both semicircular and realistic DOS are renormalized in
such a way that they run in interval [$-D,D$] with norm equal to
one. First two upper curves presented in Fig.~\ref{fig:mu_n}
correspond to $U=3D$. The upper curve obtained using
tight-binding DOS and lower one comes from semicircular DOS. The
first curve is nearly a straight line crossing $n=1$ point while
the line corresponding to semicircular DOS is about to make a
jump which is clearly presented in the behaviour of $U=4D$ line.
The jump becomes even more pronounced for $U=5D$ and both,
semicircular and tight-binding, DOSes. Let us notice that
absolute value of the jump for tight-binding DOS is smaller that
for semicircular DOS. From this figure one can easily conclude
that the critical interaction when insulating behaviour appears
in the system should be somewhere  between $U=3D$ and $U=4D$
closer to the second value (the final conclusion about the
insulating behaviour one can make from energy dependence of DOS
on real axis).

Similar to the $\mu$ against $n$ dependence the MIT behaviour one
can observe from the doping dependence of the quasiparticle
residue which is presented in Fig.~\ref{fig:z_n}. In
Fig.~\ref{fig:z_n} we plot five curves for the same values of
interaction and shapes of DOSes as in the previous graph. As we
expected for $U=3$ (both DOSes) and $n=1$ we have a finite value
of $Z$. Notice that again (as in the previous plot) tight-binding
DOS shows more metallic behaviour (larger value of $Z$ and more
straight line than in the case of semicircular DOS). All other
values of the interaction clearly show insulating behaviour of
the system.
\begin{figure}
\centering
\epsfig{file=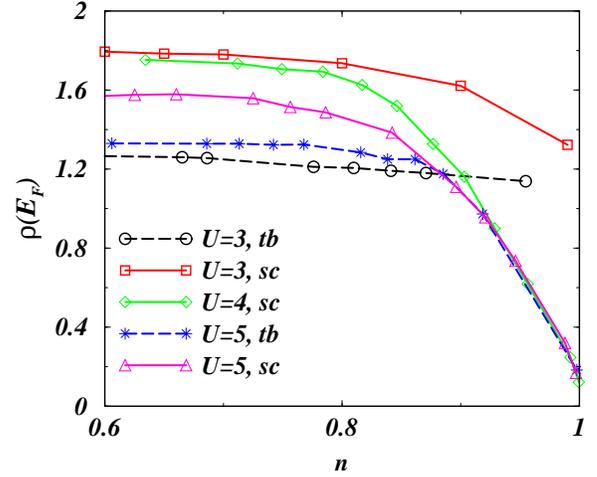,angle=0,width=3.0in}\\
\caption{DOS at the Fermi level, $\rho(E_F)$ $[states/(eV
unitcell)]$, vs filling, $n$, for semicircular (sc) and
tight-binding (tb) DOSes and various values of interaction, $U$.
All of the data was computed for $\beta = 16$.
}%
\label{fig:dos0_n}
\end{figure}

So, now when one can see how electron correlations change
physical behaviour of the system, we remind that the main input to
DMFT-QMC or DMFT-SUNCA procedure consists from the shape of DOS
(semicircle or tight-binding) and the value of interaction, $U$.
We will analyze both shapes of DOSes for two values of $U$
mentioned above.

Using our results presented in Fig.~\ref{fig:z_n} and behaviour
of $\rho(E_F)$ presented in Fig.~\ref{fig:dos0_n} we can
calculate the linear coefficient of specific heat, $\gamma$. As
we saw above, LDA results differ a lot from experimental values
for $\gamma$. Now we want to know whether we can get any
improvements applying DMFT, which changes the quasiparticle
residue, $Z$, and renormalizes DOS.

\begin{figure}[h]
\centering \epsfig{file=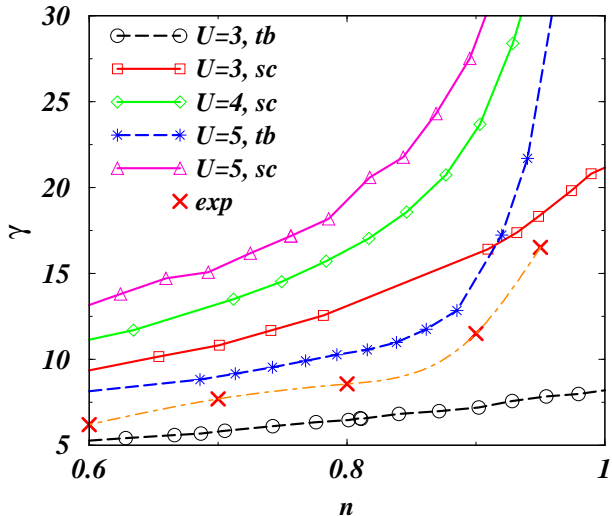,angle=0,width=3.2in}
\caption{The linear coefficient of specific heat, $\gamma$
$[mJ/mol K^{2}]$, vs. the density for different interaction
strength and DOSes at temperature $\beta = 16$.
 }%
\label{fig:dmft_gamma}
\end{figure}

In Fig.~\ref{fig:dmft_gamma} we plot the linear coefficient of
specific heat against filling for different values of Coulomb
repulsion and semicircular and tight-binding DOSes. We can notice
that for the same repulsion strength the linear coefficient of
specific heat  for the semicircular DOS is larger than for
tight-binding DOS which follows from larger pinning value in the
case of semicircular DOS. Comparing $U$ dependencies for the
semicircular DOSes we see that the linear coefficient of specific
heat for $U=3D$ is nearly linear function till filling $n=1$
which one can explain by almost linear dependence of the
quasiparticle residue. For $U=4D$ and $U=5D$ doping dependence of
the linear coefficient of specific heat reproduces the
experimental behaviour and the only question left is how close
theoretical and experimental results are.  From the plot we see
that in general results the semicircular DOSes are far from the
experiment while for the tight-binding DOS the experimental curve
just in between $U=3D$ and $U=5D$ lines. We can claim a rather
good agreement (contrary to LDA situation) between DMFT and
experimental curves for whole range of dopings. The divergence of
the linear coefficient of specific heat shows the $d$-electron
effective mass at the Fermi level is strongly enhanced on
approaching the MIT~\cite{Kumagai:1993}. In the case of $U=5D$ a
small ``overshooting" of the linear coefficient of specific heat
for large doping can be explained by 10-15~\% inaccuracy in the
procedure of the quasiparticle extraction (we define it from the
self-energy on Matsubara axis), plus one can slightly tune the
interaction strength, which probably should be smaller something
like $4.5-4.8D$.  In general agreement between DMFT-QMC results
and experimental one is quite good.

So we can summarize the linear coefficient of specific heat
results saying that changes in the spectral weight, $Z$, are the
main source of the improvement of our results for the linear
coefficient of the specific heat. Those changes are most
remarkable for $U=5D$ where $Z$ tends to zero while density
approaches the integer filling $n=1$. Diverging behaviour of
linear coefficient  of specific heat  for small doping  in the
real material can be explained  by one of the structural
transitions happening in LaTiO$_3$, at doping less than 5\% the
three-fold degeneracy is lifted and we effectively have only
one-band model for which $U=3D$ could be large enough to get the
MIT transition at integer filling.

Besides the degeneracy and the structural transition, temperature
belongs to major players on the field of the MIT. The energy
scale which separates low and high temperatures is given by $D' =
ZD$ where $D$ is uncorrelated bandwidth of the system and $Z$ is
the quasiparticle residue. Since the metal-insulator transition
is accompanied by the vanishing of the quasiparticle residue we
see that in order to access the low-temperature phase we need to
go to lower and lower temperatures as we get closer to integer
filling.  Hence we can not expect to be able to see clear
evidence of the divergence of $\gamma$ for the range of
temperatures we are working in QMC approaching integer filling.

\subsection{Comparison QMC and SUNCA}%
\label{sec:comp_qmc_sunca}

In this section we analyze and compare two impurity solvers i.e.
QMC and SUNCA which have been described in section
~\ref{sec:imsolvers}. Below we present results of those two
methods and choose one of them to compute transport. Similar to
the case of LDA calculation, in calculation the impurity problem
we also need to make a compromise between speed and accuracy. It
is well known that QMC impurity solver is very expensive but exact
(the only  approximation used in QMC is the Trotter break up)
while SUNCA is cheap method but it is based on more
approximations. QMC works in imaginary time and Matsubara
frequency domain while SUNCA works on real frequency axis. To
compute transport properties one needs the self-energy on real
axis. In the case of QMC it is necessary to make the analytical
continuation using maximum entropy or singular decomposition
method to get the self-energy on real axis as it was described in
section ~\ref{sec:acontinuation}. This is the weakest point in
the DMFT-QMC procedure. DMFT-SUNCA working on real axis has the
self-energy immediately after the self-consistency is reached.

\begin{figure}[h]
\centering \epsfig{file=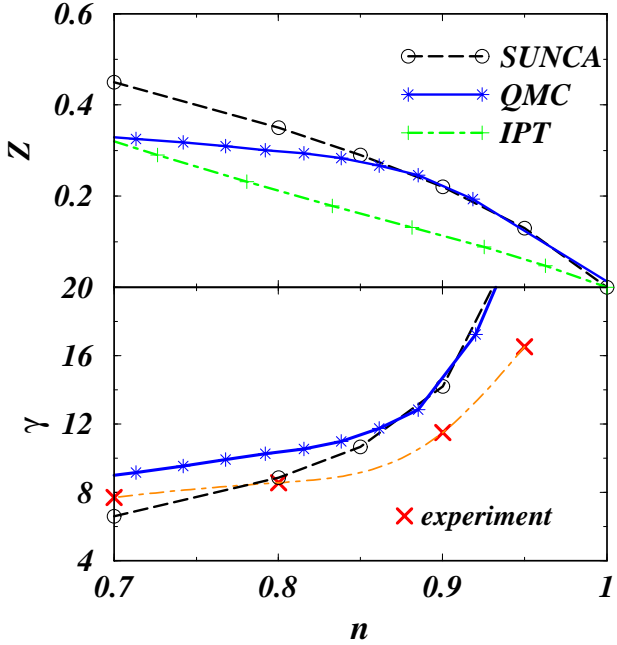,angle=0,width=3.2in}
\caption{Filling dependence of the quasiparticle residue, $Z$, and
the linear coefficient of specific heat, $\gamma$,  obtained from
two impurity solvers: QMC (solid line with stars) and SUNCA
(dashed line with circle symbols) for $U=5D$, temperature
$\beta=16$. Experimental points are given by cross symbols and
dot-dashed line is used as a guide for eye. Tight-binding density
of states was used in the self-consistency loop of the DMFT
procedure. For comparison we also provide $Z$ vs  $n$ curve
obtained with IPT method for the same parameters as ones used  in
QMC and SUNCA calculations.%
}%
\label{fig:Z_qmc_sunca}
\end{figure}

As we noticed in the previous subsection the main task of the
interaction (read the impurity solver) is to produce the MIT at
integer fillings. And one of the criteria of insulating behaviour
in the system is vanishing quasiparticle weight. In
Fig.~\ref{fig:Z_qmc_sunca}  we compare the quasiparticle residue,
$Z$ obtained from DMFT-QMC and DMFT-SUNCA methods as function of
doping for $U=5D$ and realistic DOS. We see that both methods are
in a good agreement with each other. We also provide $Z$ versus
$n$ curve calculated using iterative perturbative theory
\cite{Kajueter:1996b} (IPT) to see that all three impurity
solvers produce at least qualitatively the same trends.

Now we can go further and compare electron GF on Matsubara axis.
Imaginary axis is a natural space of work for QMC and to compare
results with SUNCA we used Lehmann representation connecting
spectral function on real axis with GF on imaginary axis. The
representation is analytical and exact, hence, the comparison can
be made without any assumptions and approximations or
uncertainties which could arise in the case of the analytical
continuation.

\begin{figure}[h]
\centering \epsfig{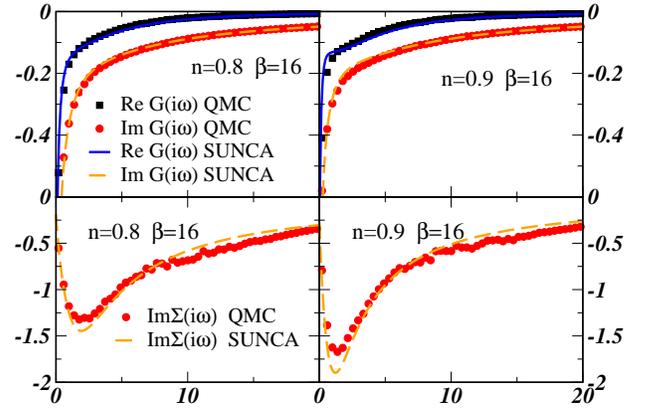}
\caption{Comparison of energy dependencies of imaginary and real
parts of GF on Matsubara axis for different dopings  computed
using two impurity solvers: QMC (circles) and SUNCA (solid line)
used in DMFT self-consistency procedure with semicircular DOS for
$U=5D$ and temperature $\beta=2$.
}%
\label{fig:comparison_b2}
\end{figure}

\begin{figure}[h]
\centering \epsfig{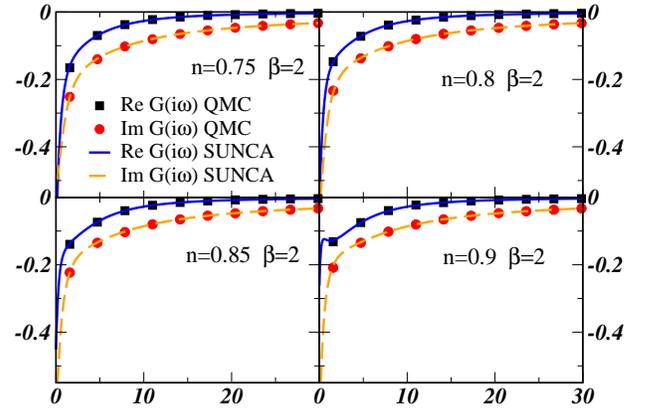}
\caption{ In the two upper panels we compare energy dependencies
of imaginary and real parts of GF on Matsubara axis for  dopings
$n=0.8$ and $0.9$ computed using QMC (circles) and SUNCA (solid
line). In the low panels we plot imaginary parts of the
self-energies for the same parameters as in the upper panels. We
used semicircular DOS in DMFT self-consistency procedure and
$U=5D$ at temperature $\beta=16$.
}%
\label{fig:comparison_b16}
\end{figure}

In Fig.~\ref{fig:comparison_b2} we plot real and imaginary parts
of GF on Matsubara axis for QMC (symbols) and SUNCA (lines) at
different dopings and temperature $\beta=2$ where one would
expect very good agreement for the methods (as higher temperature
as better and faster both methods work). And indeed it is the
case for all the curves. In Fig.~\ref{fig:comparison_b16} we plot
GF and imaginary parts of the self-energy for lower temperature,
$\beta=16$ (temperature which is mostly used in our calculations)
and for 10 and 20~\% of dopings (they are our usual dopings used
in calculations).

As we can conclude from presented curves we have quite good
agreement between the two methods and hence we can use SUNCA in
our transport calculations where behaviour of the self-energy on
real axis around the Fermi level is very crucial for the
transport properties which are extremely sensitive to the shape,
slope and value of the self-energy at the Fermi level. Transport
properties become more and more sensitive to all the details of
the transport function at the Fermi energy with lowering
temperature. Taking into account all the comparisons made and
calculations done we conclude that SUNCA is fast and accurate
enough method in comparison with QMC to compute the transport
properties of the compound.

\section{Results of transport calculations}%
\label{sec:transp_results}

\subsection{Spectral and transport functions in the real system}%
\label{sec:dos_tranfunc_for_rc}

Before to do transport computations it is worth to study spectral
and transport functions dependencies on doping and temperature. As
we discussed in the previous section ~\ref{sec:comp_qmc_sunca} we
will use SUNCA as main method to compute transport properties
(one can avoid the analytical continuation procedure in this
case). But, at any rate, we also did calculations with QMC
impurity solver and compare results obtained from the two
impurity solvers and describe differences between them when they
are the most noticeable.

\begin{figure}[h]
\centering \epsfig{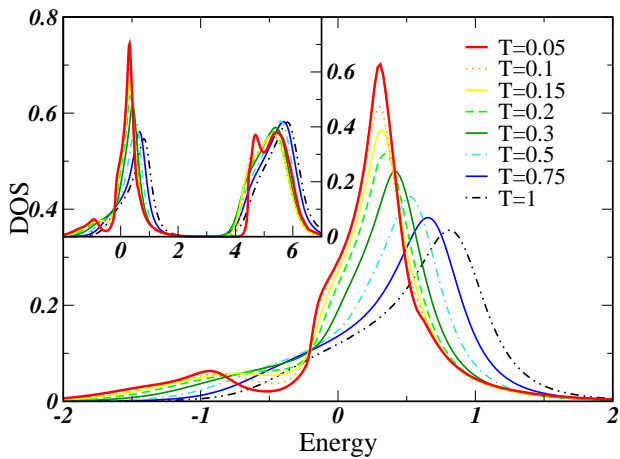}
\caption{Temperature dependence of DMFT density of states for $n
=0.8 $ and $U=5D$. Larger frequency interval is plotted in
the  insert. Energy is in units of half bandwidth, $D$.}%
\label{fig:sunca_dosesT}
\end{figure}

In Fig.~\ref{fig:sunca_dosesT} we plotted the density of states
per spin (the lower Hubbard band and quasiparticle peak are shown
in the main panel and the inset shows the whole energy range) at
filling $n = 0.8$ for various values of temperature. Here
temperature is measured in units of the half-bandwidth, $D =
1.35$~eV and thus the actual temperature range is quite large
with the smallest temperature, corresponding to $T = 0.05D$,
being around 780~K. The highest temperature plotted is equal to
one but it is still not large enough to make incoherent motion in
the system to be dominant. As we can see temperature changes are
quite substantial (the lower Hubbard band nearly disappeared and
quasiparticle peak is shifted towards the upper Hubbard band,
indicating  tendency to join the upper Hubbard band and form
incoherent broad bump) but they are still not enough to reach the
incoherent motion state (the upper Hubbard band is changed but
still it is very good separated from the QP peak -- lower Hubbard
band creation). This situation is the expected one as we know the
QP picture disappears for temperature higher then Coulomb
repulsion, $U$, which is $5D$ in our case. Hence, for $T\ge 5D$
one will see only incoherent motion in the system. Let us notice
here the difference between SUNCA and QMC where in the last
method the spectral density is just a single hump corresponding
to purely incoherent carrier dynamics observed already for
temperature $\beta=1$. If we start from incoherent picture and
lowering temperature, then the incoherent hump splits up and the
Hubbard bands start to form. For even lower temperature the lower
Hubbard band moves completely below the Fermi-surface and the
coherent quasiparticle peak appears at the Fermi-level. The lower
Hubbard band starts to form at $\beta=4$, QP peak is formed for
$\beta \ge 10$. For temperature lower then $\beta=16$ weight of
the QP peak nearly does not change. We observe similar behaviour
of DOS in SUNCA where the shape of the QP and the lower Hubbard
band change only slightly for temperatures lower then $T=0.1D$.
Described discrepancies on real axis between the two methods are
entirely in the domain of the analytical continuation which is
maximum entropy method which reliably reproduces only low-energy
part. One more interesting thing we should notice in
Fig.~\ref{fig:sunca_dosesT} which is temperature dependence of
DOS value at the Fermi level. When this value reaches the one of
non-interacting DOS we say that the pinning condition is obeyed.
The temperature when the pinning condition is reached is called
the pinning temperature and it strongly depends on doping.  For
filling $n=0.8$ as we can conclude from
Fig.~\ref{fig:sunca_dosesT} the pinning temperature is about
$0.1D$.

\begin{figure}[h]
\centering \epsfig{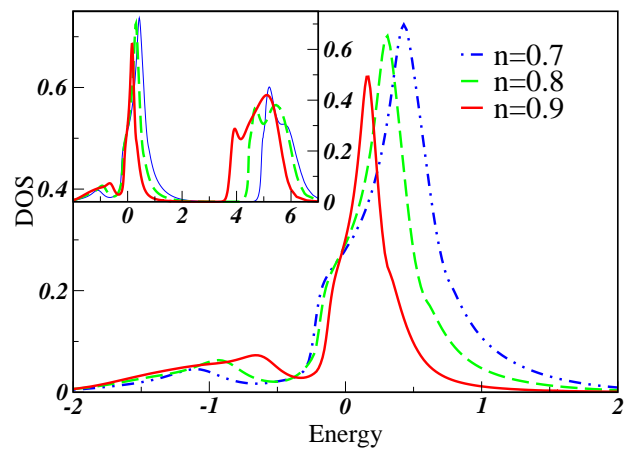}
\caption{Doping dependence of DMFT density of states for $T =
0.05 $ and $U=5D$. Larger frequency interval is plotted in
the  insert. Energy is in units of half bandwidth, $D$. }%
\label{fig:sunca_dosesn}
\end{figure}

In Fig.~\ref{fig:sunca_dosesn} we plotted the density of states
per spin for $T = 0.05$ and different values of doping. Choice of
temperature was dictated by consideration that it should be lower
than the pinning temperature for the largest filling presented.
With doping growing the quasiparticle peak broadens and its
spectral weight increases a lot while the weight of the lower
Hubbard band changes a little (doping changes are 10-20~\%). All
the weight, which the QP peak gained, came from the upper Hubbard
band (see insert in Fig.~\ref{fig:sunca_dosesn} where larger
energy interval is presented).  With doping increasing the system
becomes less and less correlated and in the limit of 100\% doping
Hubbard bands vanish and the quasiparticle peak transforms into
free and empty tight-binding band. With doping decreasing the QP
peak vanishes and the system becomes insulating for the repulsion
$U=5D$. As we mentioned it before, for three-band degenerate
Hubbard model, $U=3D$, which is often used in description of
LaTiO$_3$, is not enough to reproduce the insulating behaviour.
which appears only for $U\approx 4.5-5D$.

\begin{figure}[h]
\centering \epsfig{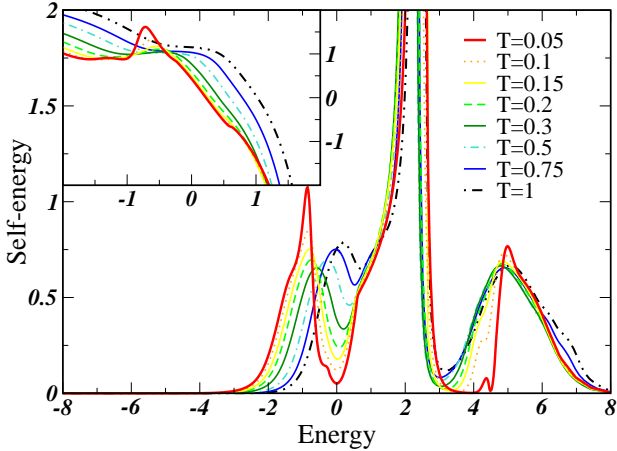}
\caption{Temperature dependence of imaginary part of the
self-energy for $n=0.8$. In the inset real part of the
self-energy is shown for the same temperatures. Energy is in
units of half bandwidth, $D$.}%
\label{fig:sunca_sigmasT}
\end{figure}

\begin{figure}[h]
\centering \epsfig{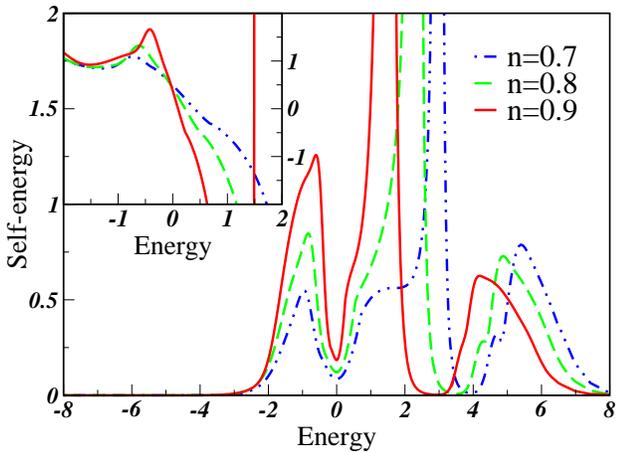}
\caption{Doping dependence of imaginary part of the self-energy
for $T=0.1$. In the inset real part of the  self-energy is shown
for the same dopings. Energy is in units of half bandwidth, $D$.}%
\label{fig:sunca_sigmasn}
\end{figure}

In Figs.~\ref{fig:sunca_sigmasT} and \ref{fig:sunca_sigmasn} we
presented dependence of imaginary part (main panels) and real part
(inserts) of the self-energy on temperature and doping  for the
same temperatures as in Figs. ~\ref{fig:sunca_sigmasT} and the
same dopings as in Fig. ~\ref{fig:sunca_sigmasn}. In Fig.
~\ref{fig:sunca_sigmasT} we see nice quadratic behavior of the
self-energy for low temperatures with the minimum at around the
chemical potential (zero in our case) which is then rises and
shifts with the temperature to the right-hand side. Real part of
the self-energy reflects the quasiparticle residue, $Z$, and with
lowering the temperature the QP residue increases and approaches
the pinning value.  The doping dependence of the imaginary part of
the self-energy shows that the self-energy at the chemical
potential decreases with increasing of doping.  This is exactly
what one should expect for a system  close to free-electron state
where more quadratic and smaller imaginary part of the
self-energy is anticipated. The real part of the self-energy shows
the same tendency with increasing doping as in the case of the
temperature dependence: the curve which crosses the Fermi level
becomes more flat. At zero doping it should have zero derivative
at the chemical potential signaling about $Z=1$. The self-energy
is extremely important characteristic of the system as it the only
quantity which enters into transport calculations. Using the
self-energy one computes the transport functions which is main
ingredient of all transport equations.

\begin{figure}[h]
\centering \epsfig{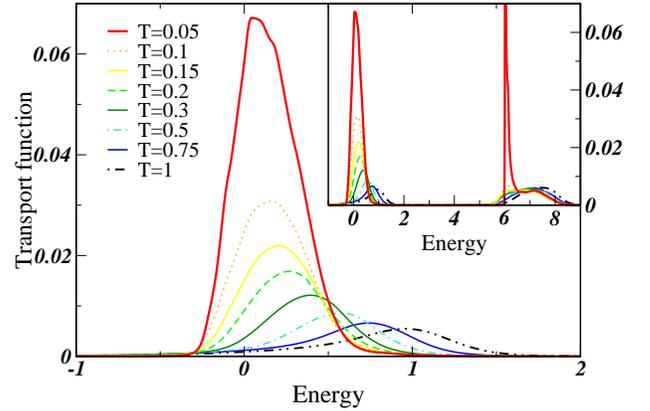}
\caption{Temperature dependence of the transport function for
$n=0.8$. In the inset larger frequency interval is used. Energy
is in units of half bandwidth, $D$.} %
\label{fig:sunca_phisT}
\end{figure}

\begin{figure}[h]
\centering \epsfig{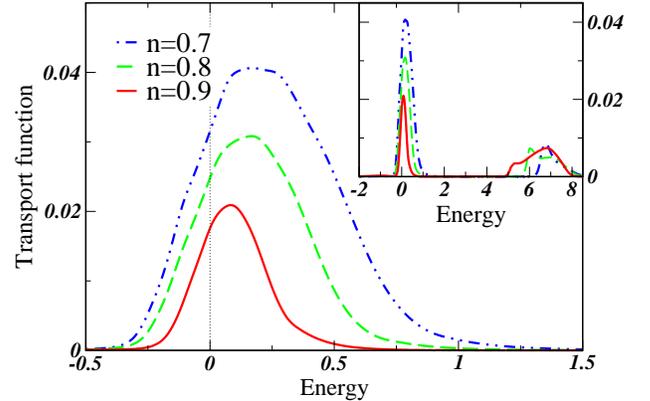}
\caption{Doping dependence of the transport function for
temperature $T=0.1$. In the inset larger frequency interval is
shown. Energy is in units of half bandwidth, $D$.}%
\label{fig:sunca_phisn}
\end{figure}

In Figs. \ref{fig:sunca_phisT} and \ref{fig:sunca_phisn} we plot
temperature and density dependencies of the transport function
for the same set of parameters as we used for Figs.~
\ref{fig:sunca_sigmasT} and \ref{fig:sunca_sigmasn},
correspondingly. One can reveal similar  features as in density
of states: in the transport function behaviour one clearly
identifies contributions coming from the upper Hubbard band and
lower one plus the QP peak. But the most important contribution
to transport properties at low temperatures comes from energy
region around the Fermi level. As it can be seen from Eqs.
~(\ref{eq:A_coefficients2}) the transport coefficients are
entirely defined by the transport function integral in an energy
window which depends on temperature. These equations allow at
least qualitatively to define at least sign of the thermopower for
small temperatures. If the slope of transport function is
uprising then the thermopower should be negative and for the
other slope it should be positive. For large energy window the
sign of the thermopower will strongly  depend on shape and
position to the chemical potential of the transport function.

\subsection{Transport parameters}%
\label{sec:transport_data}

\begin{figure}[h]
\centering \epsfig{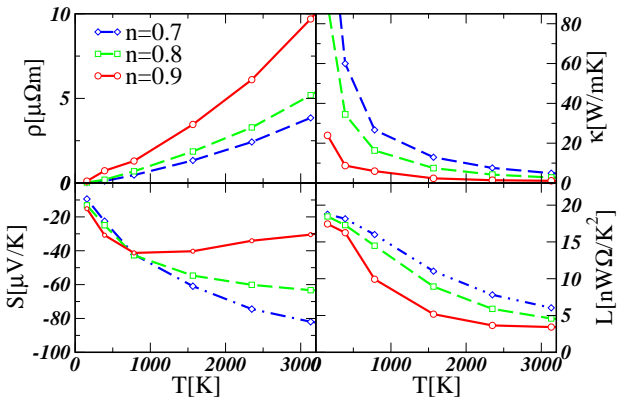}
\caption{Temperature and doping  dependencies of transport
parameters: $\rho$ denotes the electrical resistivity, $\kappa$
is the thermal conductivity, $S$ is the thermopower and $L$ is
the Lorentz ratio.  }%
\label{fig:sunca_transport1}
\end{figure}

\begin{figure}[h]
\centering \epsfig{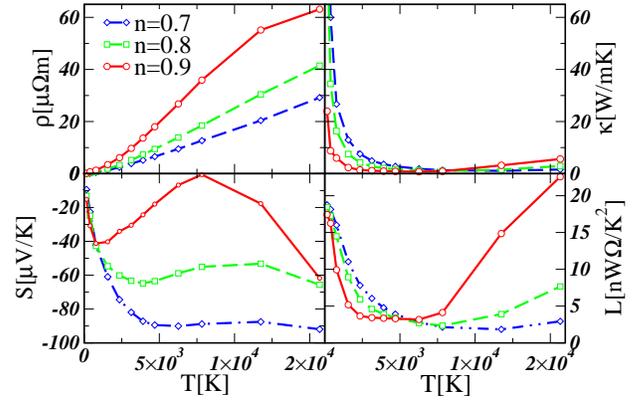}
\caption{The same transport parameters  the same as in Fig.
~\ref{fig:sunca_transport1} but on substantially larger
temperature interval. }%
\label{fig:sunca_transport}
\end{figure}

In Fig.~\ref{fig:sunca_transport1} we plotted the transport
parameters of the studied system for different densities against
temperature. The transport parameters under consideration are the
following: $\rho$ denotes the electrical resistivity, $\kappa$ is
the thermal conductivity, $S$ is the thermopower and $L$ is the
Lorentz ratio. The resistivity behaviour, as it was found
experimentally ~\cite{Tokura:1993:PRL} and theoretically, is a
quadratic function at relatively low-temperature interval
becoming linear at higher temperatures (see Fig.
~\ref{fig:sunca_transport}). The quadratic temperature dependence
of the electrical resistivity is reminiscent of the strong
electron-electron scattering which predominate the
electron--phonon scattering process. The thermal conductivity
behaves as $T^{-2}$ till temperatures of order $10^{3}-10^{4}$,
which are relatively large temperatures ~\cite{Palsson:1998}. The
Lorentz number tends to constant value around 16-17
$nW\Omega/K^2$ indicating the character of the low-temperature
scattering as Fermi-liquid one. The thermopower behaviour is
little bit more complicated. At low temperature the thermopower
linearly tends to zero. It is very hard for us to distinguish
doping dependence for relatively small temperatures as all
changes lay between error bars  which are in our case larger than
the difference between lower and higher thermopower curves
presented in the figure. The reason for large errors lays in a
very small value of imaginary part of the self-energy which we
have to deal with lowering the temperature and this situation is
very challenging for the used impurity solvers. For higher
temperatures (higher than 1000~K) we are curtain in the
thermopower behaviour as there is no any problems with the
self-energy determination in this temperature range. With
increasing temperature we observe a local maximum in the
temperature interval $5\times 10^3 - 2\times 10^4$.  We associate
it with increasing the temperature cut-off (see Eq.
~(\ref{eq:A_coefficients2})) which is large enough to take into
account right-hand side slope of the central part in the
transport function. Or in another words the local maximum in the
thermopower in some way mimics  behaviour of the transport
function (the hump around the chemical potential).

Analyzing Figs.~\ref{fig:sunca_transport1} and
\ref{fig:sunca_transport} as functions of doping for a fixed
temperature we can see that all curves behave in the way one
would expect. The resistivity is growing with decreasing doping
as the system approaches the MIT while the thermal conductivity
and the Lorentz number are decreasing.  The thermopower has a bit
more complicated behaviour which depends on fixed temperature
(which temperature slice we take). But generally it is growing
with vanishing doping and for lower than 10~\% doping one could
get even positive thermopower which first becomes positive at
temperatures around 5000~K (see $n=0.9$ thermopower curve) and
then positiveness will propagate to smaller temperatures.
Comparing our results with experimental situation in doped
LaTiO$_3$ ~\cite{Moos:1995,Hays:1999} we notice that majority of
experiments are done for temperature less than 300~K which is a
rather hard task to deal with for the reason we pointed out
above. The biggest discrepancy we found for resistivity at small
temperatures where our resistivity 3-4 times lower than the
experimental one. While the thermopower behaviour (which we also
treat as electronic one) is accurate within 30~\% in absolute
value. One would expect that the thermopower could become
positive with decreasing doping in the way it is experimentally
observed. We also could obtain it once we do a much more delicate
and hard job taking into account the structural transition
happening at doping $x<0.05$ as effectively we should have
one-band model instead of three-fold degenerate. But this is
beyond of the scope of the present work. Close to the MIT we have
strongly asymmetric DOS and transport functions which in the case
of integer filling $n=1$ will produce positive sign of the
thermopower. The reason for this is the position of negative
slope (right-hand side) of the lower Hubbard band with is closer
to the Fermi energy than the upper one and hence has dominant
contribution into the transport properties of the system.

So, it would be fair to say that  our calculations  we can catch
at least semi-qualitative behaviour of the transport parameters.
The electrical resistivity would require an additional treatment
to get quantitatively a good agreement while the thermopower
calculations deserve quantitative comparison with experiment and
can be accurate enough providing 20-30~\% agreement with
experiment.

\section{Conclusion}%
\label{sec:conclusion}

In the paper we proposed and implemented a new method for
calculation of thermoelectrical properties in real materials.
Dynamical mean-field theory was used to take into account strong
electron interactions and thereby bring the self-energy into
first-principal calculations.  Taking a rather generic for many
strongly correlated materials density of states, we obtained
temperature and doping dependencies for such thermoelectric
properties as electrical resistivity, the thermal conductivity,
the thermopower and the Lorentz ratio.

We believe that the new method will be a powerful tool for the
analysis of existing experimental data and guiding us to a proper
physical understanding of thermoelectrical phenomena.  This is
especially important not only for correlated materials such as
Mott-Hubbard insulators and high-temperature superconductors but
also for simple materials like the noble metals which display
thermoelectric behavior that still lacks a proper description. In
addition we hope this new method will aid in the search for new
materials with better thermoelectrical performance by allowing
for ab initio predictions of thermoelectric properties.

\acknowledgments

We would like to thank A.~Lichtenstein for many fruitful and
stimulation discussions. We greatly acknowledge usage of Cray
T3E-900 computer at NERSC, Berkeley, as well as Rutgers Beowulf
computational cluster which allowed us to make present
computations feasible.

\appendix

\section{Many Body Theory in a Non-orthogonal Basis}%
\label{sec:no_mbt}

Our starting point here is a representation of the kinetic term
of the Hamiltonian in an orthogonal basis, $\{|i\rangle\}$ and we
assume that this basis is related to the non-orthogonal basis,
$\{|\alpha\rangle\}$ by the transformation matrix
\begin{equation}
|i\rangle = \sum_{\alpha}|\alpha\rangle S_{\alpha i}
\qquad\mbox{and}\qquad \langle i|      =
\sum_{\alpha}\langle\alpha|S^{*}_{\alpha i} =
\sum_{\alpha}S^{+}_{i\alpha}\langle\alpha| .
\end{equation}
The Hamiltonian is now given by
\begin{eqnarray}
H       &=& \sum_{ij}\langle i|H|j\rangle c^{+}_{i}c_{j}\\\nonumber%
        &=& \sum_{ij\alpha\beta}S^{+}_{i\alpha}
           \langle\alpha|H|\beta\rangle S_{\beta j} c^{+}_{i}c_{j}
        = \sum_{\alpha\beta}H_{\alpha\beta}c^{+}_{\alpha}c_{\beta}.
\end{eqnarray}
The last term in the equation above is a requirement that we
place on the creation and destruction operators in the
non-orthogonal basis and thus we find that
\begin{equation}
c^{+}_{\alpha} = \sum_{i}c^{+}_{i}S^{+}_{i\alpha}
\qquad\mbox{and}\qquad c_{\alpha} = \sum_{j}S_{\alpha j}c_{j}.
\end{equation}
The non-orthogonality of the basis is encoded in the overlap
matrix, $O_{\alpha\beta} = \langle\alpha|\beta\rangle$ and this
matrix can be related to the transformation matrix, $S$ in the
following manner
\begin{equation}
\delta_{ij}     =  \langle i|j\rangle
                = \sum_{\alpha\beta}
        S^{+}_{i\alpha}\langle\alpha|\beta\rangle S_{\beta j}
                = \sum_{\alpha\beta}S^{+}_{i\alpha}O_{\alpha\beta}S_{\beta j}
\end{equation}
Therefore we see that the overlap matrix is given by
\begin{equation}
O = (SS^{+})^{-1}.
\end{equation}
We should note here that the creation operator $c^{+}_{\alpha}$
does not create a particle in the state $|\alpha\rangle$ when
acting on the vacuum, since as we see
\begin{eqnarray}
c^{+}_{\alpha}|0\rangle &=&
\sum_{i}c^{+}_{i}S^{+}_{i\alpha}|0\rangle = \sum_{i}|i\rangle
S^{+}_{i\alpha} \\\nonumber %
&=& \sum_{i\beta}|\beta\rangle
S_{\beta i} S^{+}_{i\alpha} = \sum_{\beta}|\beta\rangle
O^{-1}_{\beta\alpha}.
\end{eqnarray}
It is however worth noting that this state has unit overlap with
the state $|\alpha\rangle$ and zero overlap with all of the other
basis-states! The commutation relationship of these operators are
the same as for regular Fermi operators except that we get
\begin{equation}
\{c^{+}_{\alpha},c_{\beta}\} = \sum_{ij}S_{\beta
j}\{c^{+}_{i},c_{j}\}S^{+}_{i\alpha} = S_{\beta i}S^{+}_{i\alpha}
= O^{-1}_{\beta\alpha} .
\end{equation}
Let us finally obtain the expression for the Green's function in
the non-orthogonal basis
\begin{equation}
G_{\alpha\beta}(\tau) = -\langle
c_{\alpha}(\tau)c^{+}_{\beta}(0)\rangle.
\end{equation}
The easiest way to calculate this Green's function is by looking
at the Lagrangian for the system in the orthogonal basis and then
simply transform it into the non-orthogonal one. We have
(summation over repeated indices implied)
\begin{eqnarray}
{\cal L} &=& c^{+}_{i}{\partial \over \partial\tau}c_{i}
         - c^{+}_{i}H_{ij}c_{j}
        \nonumber \\
         &=& S^{-1}_{i\beta }
           c^{+}_{\alpha}{\partial \over \partial\tau}c_{\beta}
           (S^{+})^{-1}_{\alpha i}
         - c^{+}_{\alpha}H_{\alpha\beta}c_{\beta}
        \nonumber \\
         &=& c^{+}_{\alpha}O_{\alpha\beta}
         {\partial \over \partial\tau}c_{\beta}
         - c^{+}_{\alpha}H_{\alpha\beta}c_{\beta}.
\end{eqnarray}
The free Matsubara Green's function can now be obtained by
Fourier transforming the operators in the Lagrangian and then the
inverse of the Green's function, $G^{0}_{\alpha\beta}(i\omega)$,
is simply the term multiplying $c^{+}_{\alpha}c_{\beta}$.  Thus we
obtain
\begin{equation}
G^{0}_{\alpha\beta}(\omega) = \left[i\omega
O-H\right]^{-1}_{\alpha\beta}.
\end{equation}
The renormalized Green's function one gets as in the orthogonal
case by adding the self-energy to the Hamiltonian and thus
\begin{equation}
G(i\omega) = \left[i\omega O - H - \Sigma\right].
\end{equation}
We should remark here that these Green's functions do not share
the same properties as their cousins in the orthogonal bases do
and in particular the total density is not given by trace of
$G(\tau = 0^{-})$.  To see that we go back to the orthogonal
basis where we know how things work and write the density
operator as
\begin{equation}
\rho    = \sum_{i}c^{+}_{i}c_{i}
        = \sum_{i\alpha\beta}S^{-1}_{i\beta}c^{+}_{\alpha}
                             c_{\beta}(S^{+})^{-1}_{\alpha i}
        =  \sum_{\alpha\beta}O_{\alpha\beta}c^{+}_{\alpha}c_{\beta}.
\end{equation}
Thus the total density of electrons in the system is
\begin{equation}
n_{tot} = \langle\rho\rangle
        =  \sum_{\alpha\beta}O_{\alpha\beta}
           \langle c^{+}_{\alpha}c_{\beta}\rangle
        =
        \sum_{\alpha\beta}G_{\beta\alpha}(\tau=0^-)O_{\alpha\beta}.
\end{equation}
We should note in particular that this means that there seems to
be no good way of assigning a density to a particular orbital in
the non-orthogonal case.

\section{LDA Hamiltonian in Non-orthogonal Base}%
\label{sec:LMTOH}

In LDA one has to solve the known Kohn-Sham equation
\begin{equation}\label{eq:sham_eq}
(-\nabla^2+V)\Psi_{{\bf k}j}=\epsilon_{{\bf k}j}\Psi_{{\bf k}j}.
\end{equation}

The eigenfunctions $\Psi_{{\bf k}j}$ are expanded in a basis set
for example the LMTO basis $\chi_{{\bf k}}^{\alpha}({\bf r})$
which is not necessary orthogonal as
\begin{equation}\label{eq:phi}
\Psi_{{\bf k}j}=\sum_{\alpha}A_{{\bf k}j}^{\alpha} \chi_{\alpha
{\bf k}}.
\end{equation}
Substituting (\ref{eq:phi}) in (\ref{eq:sham_eq}) we obtain
$$
H_{LDA}^{\alpha\beta}({\bf k})A_{{\bf k}j}^{\beta}=\epsilon_{{\bf
k}j}O^{\alpha\beta}_{\bf k}A_{{\bf k}j}^{\beta},
$$
with the normalization condition
$$
\sum_{\alpha\beta} A_{{\bf k}j}^{\alpha *} O_{{\bf
k}}^{\alpha\beta} A_{{\bf k}j'}^{\beta}=\delta_{jj'}.
$$
The density of states with a particular character $\alpha,\beta$
is defined as
\begin{equation}\label{eq:delta_beta}
\rho^{\alpha,\beta}(\epsilon)=\sum_{{\bf k}j} A_{{\bf k}j}^{\alpha
*} O_{{\bf k}}^{\alpha\beta} A_{{\bf
k}j}^{\beta}\delta(\epsilon-\epsilon_{{\bf k}j}).
\end{equation}

The DOS of the tight-binding fit one can get from
Eq.~(\ref{eq:delta_beta}) restricting index  $j$ to the three
bands near the Fermi level. Fig. \ref{fig:realdoses} displays the
partial $t_{2g}$ DOS of LaTiO$_3$ (Eq.~(\ref{eq:delta_beta}) with
$\alpha$ of $t_{2g}$ character). The total DOS one can get from
Eq.~(\ref{eq:delta_beta}) summing over indexes $\alpha,\beta$
\begin{equation}\label{eq:tot_dos}
\rho(\epsilon) = \sum_{{\bf k}j} \delta ( \epsilon-\epsilon_{{\bf
k}j}).
\end{equation}

Notice that the eigenvectors $A_{{\bf k}j}^{\alpha}$ are the
matrix elements ($S_{{\bf k}j}^{\alpha}=A_{{\bf k}j}^{\alpha}$)
of the Cholesky decomposition of $O^{-1}$ defined in
Appendix~\ref{sec:no_mbt} and that the partial DOS is normalized
to be of one integrated over the whole frequency range. Held et
al.~\cite{Held:2001} proposed to use the partial $t_{2g}$ DOS, the
low-frequency range with a recalling so as to normalize it to one
\begin{equation}\label{eq:n_t2g}
n_{t_{2g} }  = \sum\limits_{\alpha  \in t_{2g} } {} \int {} \rho
_{}^{\alpha \alpha } (\varepsilon )f(\varepsilon ).
\end{equation}

\section{Splines and Fourier Transformations}%
\label{app:spline}

\subsection{Direct Fourier Transformation}%
\label{app:directFT}

In QMC program we do the direct Fourier transformation exactly,
i.e. first we obtain coefficients of the cubic spline exploiting
physical properties of GF transformed and then make analytical
Fourier integration knowing the form of the splined curve. The
cubic spline interpolation formula reads
\[
G(\tau )=a_{i}+b_{i}(\tau -\tau _{i})+c_{i}(\tau -\tau
_{i})^{2}+d_{i}(\tau -\tau _{i})^{3}, \;
\tau\in[\tau_i,\tau_{i+1}],
\]
where coefficients $a_{i}, b_i, c_i, d_i$ are equal to values of
the function, its first, second and third derivatives at knot $i$
i.e. $a_{i}=G(\tau _{i}),\,b_{i}=G^{\prime }(\tau
_{i}),\,c_{i}=G^{\prime \prime }(\tau _{i}),\,d_{i}=G^{\prime
\prime \prime }(\tau _{i}).$

Or in terms of GF values, $G_{i}=G(\tau _{i}),$ and its second
derivative, $M_{i}=G^{\prime \prime }(\tau _{i})$, only
\begin{eqnarray}
a_{i} &=&G_{i},  \label{ABCDcoefficients} \\
b_{i} &=&\frac{G_{i+1}-G_{i}}{h}-\frac{2M_{i}+M_{i+1}}{6}h,  \nonumber \\
c_{i} &=&\frac{M_{i}}{2},  \nonumber \\
d_{i} &=&\frac{M_{i+1}-M_{i}}{6h}.  \nonumber
\end{eqnarray}

From equations above we see that one needs to know  the second
derivatives, $M_i$, using tabulated values of GF, $G_i$, in order
to get the cubic spline interpolation. To obtain $M_{i}$
coefficients we use conditions of smoothness of the first
derivative and continuity of the second one. As the result we
have $L+1$ equations for $L+3$ unknowns
\begin{equation}
\left[
\begin{array}{cccccc}
2 & \lambda _{0} &  &  &  & 0 \\
\mu _{1} & 2 & \lambda _{1} &  &  &  \\
& \mu _{2} & . & . &  &  \\
&  & . & . & . &  \\
&  & . & . & 2 & \mu _{n-1} \\
0 &  &  &  & \mu _{n} & 2
\end{array}
\right] \left[
\begin{array}{c}
M_{0} \\
M_{1} \\
... \\
... \\
\\
M_{n}
\end{array}
\right] =\left[
\begin{array}{c}
d_{0} \\
d_{1} \\
. \\
. \\
. \\
d_{n}
\end{array}
\right] \label{SYTEMOFEQUATIONS}
\end{equation}
where $L$ is a number of time slices. In addition to $L+1$ $M_0,
... M_n$, $n=0,...L$, unknowns $d_0$ and $d_n$ also should be
provided. Last two unknowns entirely depend on the boundary
conditions which we have to specify in order to have a unique
solution of Eq.~(\ref{SYTEMOFEQUATIONS}). If one knows the first
derivatives at the end-points then $d_0$ and $d_n$ are defined
through
\begin{eqnarray*}
\lambda _{0} &=&1,\,\ d_{0}=\frac{6}{h}\left( \frac{G_{1}-G_{0}}{h}%
-G_{0}^{\prime }\right) ,\\
\mu _{0} &=&1,\ d_{n}=\frac{6}{h}\left( G_{n}^{\prime
}-\frac{G_{n}-G_{n-1}}{h}\right),\\
\end{eqnarray*}
and $ d_{i}=\frac{3}{h}\left(
\frac{G_{i+1}-G_{i}}{h}-\frac{G_{i}-G_{i-1}}{h} \right),$ %
$\lambda _{i}=\mu _{i}=\frac{1}{2},$ for $i\in \lbrack 1,n-1].$
More detailed derivations of the above formulae one can find in
Ref.~\onlinecite{Stoer:1980}.

We can reduce number of unknowns just putting $M_0$ and $M_n$ to
zero (it is so called natural spline boundary conditions). In
this case
$$
\lambda _{0}=0,\,d_{0}=0,\mu _{n}=0,d_{n}=0,
$$
and we have the number of unknowns matching the number of
equations, $L+1$.

This boundary condition is good enough to compute FT of GF in the
system at or close to half filling since the second derivative of
the Green's function is small in absolute value in this regime.
And using the natural spline boundary condition we do not impose a
noticeable error. However, away from half filling when the
asymmetry of the system grows, along with amplitude, of one out of
the two second derivatives, usage of the natural spline
eventually leads to pathological behavior of the self-energy. The
signature of this pathology is in the ``overshooting" effect
~\cite{Oudovenko:2002} when the self-energy at some finite
Matsubara frequency i.e. the imaginary part of the self-energy,
becomes positive in some frequency region on the positive
Matsubara half-axis while it should be always negative. This, of
course, amounts to having negative spectral weight for the
self-energy which is something that does not occur for fermionic
response functions. The ``overshooting" can get especially severe
in the limiting cases of small temperatures, small particle
densities or large interaction strength.

So, to avoid the problem with the self-energy and, hence, with the
whole procedure of the self-consistency in DMFT-QMC program we
need to use the proper boundary conditions. And in this case we
have two possibilities to get unique solution for the system of
Eq.~(\ref{SYTEMOFEQUATIONS}) exploiting physical properties of
studied GF: a) we can provide the first derivatives at both ends
separately (in the next section we show how to calculate those
derivatives) or b) we can provide the sum of the first and the
sum of the second derivatives at the end-points, so called the
first and the second moments of GF.

With the second choice of the boundary conditions (b) the system
of equations become three-diagonal one with two off-diagonal
elements in the opposite corners of the matrix ($-M_{n-1}$ and
$-\frac{1}{2}M_{0} $)
\begin{equation}
\begin{array}{rrrllll}
4M_{0} & +\phantom{1}M_{1} & &  & &-M_{n-1} & =d_{0} \\%
\frac{1}{2}M_{0} & +2M_{1} & +\frac{1}{2}M_{2}& & & &=d_{1} \\%
&\frac{1}{2}M_{1}&+2M_{2} & +\frac{1}{2}M_{3}  & & &=d_{2} \\ %
& & \frac{1}{2}M_{2} & +2M_{3} & +\frac{1}{2}M_{4}& &=d_{3} \\ %
& & & \ddots  & \ddots  & \ddots & \vdots  \\ %
& & & \frac{1}{2}M_{n-3}& +2M_{n-2}& \frac{1}{2}M_{n-1}&=d_{n-2} \\%
-\frac{1}{2}M_{0} & & & & +\frac{1}{2}M_{n-2} & +2M_{n-1}&=d_{n-1} %
\end{array}
\label{SYTEMOFEQUATIONS1}
\end{equation}
where
$d_0=\frac{6}{h}\left( {\frac{{G_1-G_0 }}{h}+%
\frac{{G_n-G_{n-1}}}{h}-M^{(1)}}\right) +2M^{(2)} $,
$d_{n-1}=\frac{6}{h}\left( {\frac{%
{G_{n}+G_{n-2}-2G_{n-1}}}{h}}\right) -\frac{1}{2}B$,
$G_{0}^{\prime }+G_{n}^{\prime }=M^{(1)}$, $M_{0}+M_{n}=M^{(2)}$.

Now having the second derivatives $M_i$ and, hence coefficients
$a_i, b_i, c_i, d_i$ we can take Fourier integral analytically
\noindent
\begin{eqnarray}
\label{mathematica} %
&&G_{m}(\omega_n) =\\\nonumber%
&& \int_{\tau_{m-1} }^{\tau_{m}}\!\!\!\!\!\!\!d\tau[ a+b(\tau - \tau_m)+c(\tau -%
\tau_m)^ 2-d(\tau - \tau_m)^3)] {\rm e}^{(i\tau\omega_n)}=  \\\nonumber %
&&\frac{{{\rm e }^{i \tau_m  \omega_n}}  (-6 d +2 i  c  \omega_n+b %
{\omega^{2}}-i a {\omega_{n}^{3}})}{{\omega_{n}^{4}}}-\\\nonumber %
&&\frac{1}{{\omega_{n}^4}} ({{\rm e }^{i  \tau_{m-1}\omega_n}}
(-6 d+2 i c \omega_n -6 i \Delta\tau d \omega_n +b {\omega_n^2}-2c %
\Delta\tau {\omega_n^2}+ \\\nonumber %
&&3 {(\Delta\tau)^2} d {\omega_n^2}-i a {\omega_n^3}+i  b
\Delta\tau {\omega_n^3}-i c {(\Delta\tau)^2} {\omega_n^3}+i
{(\Delta\tau)^3} d {\omega_n^3})).\\\nonumber
\end{eqnarray}

Sum $G_{m}(\omega_n)$ over $m$
$$
G(\omega_n) =\sum_{m=1}^{L}G_{m}(\omega_n),
$$
will give us the Fourier integral in frequency space.

\subsection{Inverse Fourier transformation}%

As it is well known Green's function $G(\omega )$ falls off as
$1/\omega $ when $\omega \rightarrow \infty $. In the program we
deal with finite number of frequency points and cutting off
$1/\omega $ tail one would make a rather crude approximation as
the discontinuity of GF $G(\tau )$ (imaginary time domain!) has
been removed. In such situation, the high-frequency tail has to
be extracted from GF $G(\omega )$ and Fourier transformed
analytically  using the following Fourier relation
\begin{equation}
\frac{1}{i\omega _{n}-\epsilon }\leftrightarrow -[\Theta (\tau
)+\zeta n(\epsilon )]\,e^{-\epsilon \tau }, %
\label{eq:tail}
\end{equation}
where $n(\epsilon )\equiv 1/[\exp \{\beta \epsilon \}-\zeta ]$
and $\zeta =\pm 1$ depending on whether $\omega _{n}$ is bosonic
or fermionic.

The inverse Fourier transformation for GF without the tail is made
by straightforward summation over Matsubara frequencies. Once it
has been done we add the information about the tail using
Eq.~(\ref{eq:tail}).

\section{Moments}%
\label{app:moments}

Moments, $M^{(k)}$, are nothing else as the expansion of GF in
frequency domain
\begin{equation}
G(\omega )=\sum\limits_{k=0}^{N} \frac{M^{(k)}}{\omega ^{k+1}}.
\label{Mi}
\end{equation}
Another definition of $k$-degree moment is the following
\begin{equation}
M^{(k)} = \int\limits_{-\infty}^{+\infty} d\omega \omega^{k} {\rho}%
_{}(\omega) ,  \label{eq:Mn}
\end{equation}
where ${\rho}_{}(\omega)$ is  density of states (DOS).

Moments $M^{(k)}$ can be bind to sum of GFs and sum of its
derivatives in imaginary-time space as
\begin{equation}  \label{eq:GM}
(-1)^{k+1}(G^{(k)}(0^{+})+G^{(k)}(\beta^{-}))=M^{(k)} ,
\end{equation}
where $k=0, \ldots N$.

To show this one needs to take Fourier integral in parts
\begin{eqnarray}
G(i\omega_n)&=&  \int\limits_{0}^{\beta} {\rm {e}}^{i\omega_{n}\tau}%
  G(\tau) d\tau \\\nonumber%
&=&\sum_{k=0}^{N} \frac{(-1)^{k+1}(G^{(k)}(0^{+})+G^{(k)}(\beta^{-}))}{%
(i\omega_n)^{k+1}} \\
&+&\frac{(-1)^{N+1}}{(i\omega_n)^{N+1}} \int\limits_{0}^{\beta} {\rm {e}}%
^{i\omega_{n}\tau} \frac{\partial ^{N+1} G(\tau)}{\partial
\tau^{N+1} } d\tau .  \nonumber  \label{eq:GFF}
\end{eqnarray}

So, to solve the system of Eq.~(\ref{SYTEMOFEQUATIONS1}) we need
to adhere to the proper boundary conditions which are expressed
through the various moments of the Green's function. What we need
finally it to provide the first three moments $M^{(0)}, M^{(1)},
M^{(2)}$.  The first moment for Green's function is equal to one,
the second moment proportional to the chemical potential in the
system and the third one is a little bit more complicated and
contains a density-density correlator. To show that we start with
the single impurity Anderson model which reads
\begin{eqnarray}\nonumber
\label{eq:AIM} H_{SIAM} &=&
 \sum\limits_{k\alpha} {} \varepsilon _{k\alpha} c_{k\alpha}^\dag c_{k\alpha}
+\sum\limits_{\alpha} (\varepsilon _{\alpha} +
\frac{1}{2}\sum_{\alpha'\ne\alpha}U_{\alpha'\alpha})f_\alpha^\dag
f_{\alpha}
\\ &+&
 \sum\limits_{k\alpha} {}
V_{k\alpha} (f_{\alpha}^\dag c_{k\alpha} + c_{k\alpha }^\dag
f_{\alpha} )
\\&+&
\sum\limits_{\alpha <}\!\sum_{\alpha^{\prime}}%
U_{\alpha\alpha'}(n_{\alpha} n_{\alpha^{\prime}} -
\frac{1}{2}(n_{\alpha} + n_{\alpha^{\prime}} )),\nonumber
\end{eqnarray}
where $\tilde \varepsilon _{\alpha}=\varepsilon _{\alpha} +
\frac{1}{2}\sum_{\alpha\ne\alpha'}U_{\alpha'\alpha} $, the first
three moments be obtained from the following commutators
\[
M^{(k)}=\left\langle \{{\cal L}^{k}f_{\alpha };f_{\alpha
}^{\dagger }\}_{+}\right\rangle ,
\]
where ${\cal LO=[O{\tt ,}H]\,}$ denotes the commutator of operator
${\cal O}$ with the Hamiltonian, and $\{...\}_{+}$ is the
anticommutator. After some algebra one finds the following
expressions for the moments
\begin{eqnarray}  \label{GFMOM}
M^{(0)} &=& \left\langle {\left\{ {f_\alpha ,f_\alpha^\dag }
\right\}} \right\rangle =
1, \\
M^{(1)} &=& \left\langle {\left\{ {[f_\alpha ,H],f_\alpha^\dag }
\right\}} \right\rangle = {\tilde \varepsilon _\alpha +
\sum\limits_{\alpha' \ne \alpha} {}
U_{\alpha\alpha'}(n_{\alpha^{\prime}} - \frac{1}{2})},  \nonumber \\
M^{(2)} &=& \left\langle {\left\{ {[[f_\alpha
,H],H],f_\alpha^\dag } \right\}} \right\rangle = \left\langle
{\left\{ {[f_\alpha ,H],[H,f_\alpha^\dag ]} \right\}}
\right\rangle =  \nonumber \\
&&\left\langle {\tilde \varepsilon _\alpha^2 + 2\tilde
\varepsilon _\alpha \sum\limits_{\alpha' \ne \alpha}
U_{\alpha\alpha'} (n_{\alpha^{\prime}} - \frac{1}{2}) + } \right.
\nonumber \\
&&\left. {}\sum\limits_{\alpha',\alpha''}\!\!\sum_{\ne\alpha}%
U_{\alpha\alpha'}U_{\alpha\alpha''}
(n_{\alpha^{\prime}}-{\frac{1}{2}}%
)(n_{\alpha^{\prime\prime}}-{\frac{1}{2}}) {\ + \sum\limits_k {}
V_{k\alpha}^2 } \right\rangle ,\nonumber
\end{eqnarray}
where $ \sum\limits_k  V_{k\alpha}^2  = M_{0}^{2}-(M_{0}^{1})^2
$, and the moments $M_0^{i}$ are defined by Eq.~(\ref{eq:Mn}) with
$\rho(\omega)= D(\omega)$, $D(\omega)$ is non-interacting DOS.

Summing up similar terms in $SU(N)$ approximation we get
\begin{eqnarray}  \label{GFSUNMOM}
M^{(1)} &=& {\varepsilon_\alpha + (2N - 1)Un} , \\
M^{(2)} &=& \varepsilon_\alpha^2 + 2\varepsilon_\alpha (2N - 1)Un +   \\
&& U^2 \left[(2N - 1) n +  \left\langle {nn} \right\rangle
 \right] + \sum\limits_k V_{k\alpha}^2 ,\nonumber
\end{eqnarray}
where $n$ is filling per band and per spin,
$n=\frac{1}{2N}\sum_{\alpha}n_{\alpha}$,  and double occupancy is
defined as $\langle {nn} \rangle =\sum\limits_{\alpha\ne \alpha'}
\langle {n_{\alpha}n_{\alpha'}} \rangle $.

The second way to make the correct cubic spline as we mentioned
before in section \ref{app:directFT} is to provide the first
derivatives at both ends of imaginary time interval (the boundary
conditions). To find the first derivatives at the ends one can use
the following definition of the first derivatives of
finite-temperature GF
\[
-\frac{\partial }{{\partial \tau }}\left\langle
{T_{\tau}f_{\alpha } (\tau )f_{\alpha }^\dag  (0)} \right\rangle
=  - \left\langle {T[H,f_{\alpha } ]f_{\alpha }^\dag  }
\right\rangle = G'_{\alpha}(0^{+} ).
\]
Using as the Hamiltonian $H = H_{SIAM}$ we can easily obtain the
derivatives at the ends
\begin{eqnarray}  \nonumber
G'_{\alpha}(0^+ )&=& \varepsilon _\alpha (1 - n_\alpha) +
\left\langle {\sum\limits_k {} V_{k\alpha} c_{k\alpha}
f_\alpha^\dag  }
\right\rangle  +\\   \nonumber%
&& \sum\limits_{\alpha'\ne \alpha}U_{\alpha\alpha'}\left(
{n_{\alpha'} - \left \langle {n_{\alpha'} n_\alpha
} \right\rangle } \right),\\\nonumber%
G'_{\alpha}(\beta^{-}) &=& \varepsilon _\alpha n_\alpha +
\left\langle {\sum\limits_k {}
V_{k\alpha} f_\alpha^\dag  c_{k\alpha}   } \right\rangle  +\\%
&& \sum\limits_{\alpha'\ne \alpha}U_{\alpha\alpha'}\left\langle
{n_{\alpha'} n_\alpha} \right\rangle, %
\label{Gderiv1}
\end{eqnarray}
where averages e.g. $\left\langle {\sum\limits_k {} V_{k\alpha}
c_{k\alpha} f_\alpha^\dag  } \right\rangle $ can be calculated
from the following expression
\[
\left\langle {\sum\limits_k {} V_{k\alpha} c_{k\alpha}
f_\alpha^\dag  } \right\rangle  = -T\sum\limits_n {} \Delta
_\alpha (i\omega _n )G_\alpha (i\omega _n ).
\]

In obtained formulae (Eqs.~(\ref{GFMOM})-(\ref{Gderiv1})\;) we
should know filling, $n_{\alpha}$, for each band and spin as well
as we should know density-density correlator $\left\langle
{n_{\alpha} n_{\alpha'}} \right\rangle$. The filling we can
extract from GF itself. Calculation of the correlator in QMC
highlights one of the advantages of the method: the correlator is
provided by the QMC itself and one does not need to rely on any
additional approximations to obtain it as e.g. in the case of
multiband IPT method \cite{Kajueter:thesis} where coherent
potential approximation is used to get the correlator. At each
time slice the density-density correlator is also computed from
GF but in imaginary time domain where it is simply a product of
two Green's functions in $(\tau, \tau')$ space. We should note
here that we compute the correlator along with other parameters
in the system at each iteration step and once the
self-consistency is reached we have correctly obtained all the
components and parameters in the system. And finally, with small
enough imaginary time step $\Delta \tau $ one can completely
avoid the ``overshooting" problem. Keeping in mind the main
limitation of QMC procedure $U\Delta \tau /2<1$. In the present
computations we choose $\Delta \tau =1/4$ which is good enough
for the range of parameters we use in the current paper.

\section{SUNCA Equations}%
\label{app:SUNCA_EQ}

In this Appendix, we explicitly give the SUNCA equations for
degenerate Anderson impurity model with $N/2$ bands considering
fluctuations between states with $M-1$, $M$ and $M+1$ electrons
on the impurity. Self-energies are analytically continued to real
frequencies and projected onto the physical $Q=1$ subspace. We
first define the ladder vertex functions $T_a$, $T_b$ with rungs
of pseudo-particles $a$ (with $M+1$ electrons) and $b$ (with
$M-1$ electrons), respectively, as shown diagrammatically in
Fig.~\ref{fig:tmatrix}. These vertex functions, obey the following
Bethe--Salpeter equations,
\begin{widetext}
\begin{eqnarray}
T_{a}(\omega,\Omega) &=& 1 + (N-M)\int{{d\epsilon}
f(\epsilon-\Omega) A^0_{c}(\epsilon-\Omega) G_{f}(\epsilon)
G_a(\epsilon+\omega-\Omega) T_{a}(\epsilon,\Omega)},\\
T_{b}(\omega,\Omega) &=& 1 + M \int{{d\epsilon}
f(\epsilon-\Omega) A^0_{c}(\Omega-\epsilon) G_{f}(\epsilon)
G_b(\epsilon+\omega-\Omega) T_{b}(\epsilon,\Omega)},
\end{eqnarray}
where $f(\epsilon)$ is the Fermi function and
$A^0_{c}(\epsilon)=-{1 \over \pi} {\rm Im} G_{c}^0(\epsilon)$ is
the bare conduction electron density of states. For concreteness,
all propagators are to be understood as the retarded ones.  The
auxiliary particle self--energies (Fig.~\ref{fig:selfen}) are
then given by,
\begin{eqnarray}
\Sigma_{f}(\omega) &=&
  M \int{{d\epsilon} f(\epsilon-\omega) A^0_{c}(\omega-\epsilon) G_b(\epsilon)
  T_{a}(\omega,\epsilon)^2}+(N-M)\int{{d\epsilon} f(\epsilon-\omega)
  A^0_{c}(\epsilon -\omega) G_a(\epsilon) T_{b}(\omega,\epsilon)^2}
\nonumber\\
  &-& 2 M(N-M)\int{{d\epsilon} f(\epsilon-\omega)
  A^0_{c}(\omega-\epsilon) G_b(\epsilon)} {
  \int{{d\epsilon^\prime}
  f(\epsilon^\prime-\epsilon) A^0_{c}(\epsilon^\prime-\epsilon)
  G_{f}(\epsilon^\prime)G_a(\epsilon^\prime+\omega-\epsilon)}},
\\
  \Sigma_b(\omega) &=& (N-M+1)\left. \int{{d\epsilon}
  f(\epsilon-\omega) A^0_{c}(\epsilon-\omega) G_{f}(\epsilon)
  T_{a}(\epsilon,\omega)} + (N-M+1)(N-M)\int{{d\epsilon}
  f(\epsilon-\omega) A^0_{c}(\epsilon-\omega) G_{f}(\epsilon)}
  \times\right.
\nonumber\\
&& \left.\int{d\epsilon^\prime}
  f(\epsilon^\prime-\omega) A^0_{c}(\epsilon^\prime-\omega)
  G_{f}(\epsilon^\prime)G_a(\epsilon^\prime+\epsilon-\omega)
  \left[T_{b}(\epsilon,\epsilon^\prime+\epsilon-\omega)\;
  T_{b}(\epsilon^\prime,\epsilon^\prime+\epsilon-\omega)-1\right]
  \right. ,
\\
  \Sigma_a(\omega) &=& (M+1) \left. \int{{d\epsilon}
  f(\epsilon-\omega) A^0_{c}(\omega-\epsilon) G_{f}(\epsilon)
  T_{b}(\epsilon,\omega)} + (M+1)M\int{{d\epsilon} f(\epsilon-\omega)
  A^0_{c}(\omega-\epsilon) G_{f}(\epsilon)}\times \right.
\nonumber\\
  && \left.\int{d\epsilon^\prime} f(\epsilon^\prime-\omega)
  A^0_{c}(\omega-\epsilon^\prime)
  G_{f}(\epsilon^\prime)G_b(\epsilon^\prime+\epsilon-\omega)
  \left[T_{a}(\epsilon,\epsilon^\prime+\epsilon-\omega)\;
  T_{a}(\epsilon^\prime,\epsilon^\prime+\epsilon-\omega)-1\right]\right.
  .
\end{eqnarray}

In order to calculate the local electron spectral function
$A_{d}$ from the self-consistently determined $G_a$, $G_b$, $G_f$,
it is convenient to define modified vertex functions as
\begin{eqnarray}
S^R_{a}(\omega,\Omega) &=& 1 + (N-M)\int{{d\epsilon}
  f(\epsilon-\Omega) A^0_{c}(\epsilon-\Omega){\rm Re} \{
  G_{f} (\epsilon)T_{a}(\epsilon,\Omega)\}
  G_a(\epsilon+\omega)},\\
S^I_{a}(\omega,\Omega) &=& (N-M)\int{{d\epsilon}
  f(\epsilon-\Omega) A^0_{c}(\epsilon-\Omega){\rm Im} \{
  G_{f} (\epsilon)T_{a}(\epsilon,\Omega)\}
  G_a(\epsilon+\omega)},\\
S^R_{b}(\omega,\Omega) &=& 1 +M \int{{d\epsilon}
  f(\epsilon-\Omega) A^0_{c}(\Omega-\epsilon) {\rm Re}\{
  G_{f}(\epsilon)T_{b}(\epsilon,\Omega)\}
  G_b(\epsilon-\omega)},\\
S^I_{b}(\omega,\Omega) &=& M \int{{d\epsilon}
  f(\epsilon-\Omega) A^0_{c}(\Omega-\epsilon) \; {\rm Im}\{
  G_{f}(\epsilon)T_{b}(\epsilon,\Omega)\}
  G_b(\epsilon-\omega)}.
\end{eqnarray}
The local spectral function then reads
\begin{eqnarray}
A_{d}(\omega) =
&-&\left(\begin{array}{c}N-1\\M-1\end{array}\right){\rm Im}
    \int \frac{d\Omega}{\pi^2} \frac{{\rm e}^{-\beta\Omega}}{f(-\omega)}
    G_{f}(\Omega+\omega)\left\{{\rm Im} [G_b(\Omega)]
    [S^R_{a}(\omega,\Omega)^2-S^I_{a}(\omega,\Omega)^2]+\right.
    \left.
    2 {\rm Re} [G_b(\Omega)]
    S^R_{a}(\omega,\Omega) S^I_{a}(\omega,\Omega)\right\}
\nonumber\\
   &-&\left(\begin{array}{c}N-1\\M\end{array}\right) {\rm Im} \int \frac{d\Omega}{\pi^2}\;
    \frac{{\rm e}^{-\beta\Omega}} {f(\omega)}\;
    G_{f}(\Omega-\omega) \left\{ {\rm Im} [G_a(\Omega)]
    [S^R_{b}(\omega,\Omega)^2-S^I_{b}(\omega,\Omega)^2]+\right.
    \left.
    2 {\rm Re} [G_a(\Omega)]
    S^R_{b}(\omega,\Omega) S^I_{b}(\omega,\Omega) \right\}
\nonumber\\
   &+&2 M \left(\begin{array}{c}N-1\\M\end{array}\right)\int\frac{d\Omega}{\pi^2} \;
    \frac{{\rm e}^{-\beta\Omega}}{f(\omega)}\;
    \int{d\epsilon} f(\epsilon-\Omega)
    A^0_{c}(\epsilon-\Omega) \;
    {\rm Im} [G_b(\Omega) G_{f}(\epsilon)]
    {\rm Im} [G_{f}(\Omega+\omega) G_a(\epsilon+\omega)].
\label{Ad}
\end{eqnarray}
\end{widetext}

Note that the exponential divergencies of the statistical factors
appearing in Eq.~(\ref{Ad}) are compensated by the threshold
behavior of the corresponding auxiliary particle spectral
functions $A_{\mu}(\omega ) = -\frac{1}{\pi} {\rm Im}
G_{\mu}(\omega)$, $\mu =a,b,f$ in the integrands.  For the
numerical treatment, these divergencies can be explicitly
absorbed by formulating the self--consistency equations
(A1)--(A10) in terms of the functions $\tilde A_{\mu}(\omega)$
which are defined via
\begin{equation}
A_{\mu}(\omega ) = f(-\omega) \tilde A_{\mu}(\omega),
\end{equation}
and, hence, have no exponential divergence.  We thus have, e.g.,
$\exp (-\beta\omega) A_{\mu}(\omega ) = f (\omega ) \tilde
A_{\mu}(\omega)$.

\section{Transport Calculations: Currents Derivation}%
\label{app:transport}

Below we derive the expressions for the currents in a general
basis. This is done by extending the gauge-theoretic method
developed in Ref.~\onlinecite{Moreno:1996}. In the non-orthogonal
basis the action for the system  can be expressed as follows
\begin{equation}
S = \int d\tau \sum_{k} c_{k\alpha}^{+} \left(
O_{k\alpha\beta}\lrpart{\tau}+ H_{k\alpha\beta} \right)c_{k\beta}.
\end{equation}

\noindent Here $\lrpart{\tau} = {1 \over
2}(\rpart{\tau}-\lpart{\tau})$ denotes the anti-symmetrized time
derivative. The particle and heat currents can now be obtained by
considering the invariance of the action under local phase
transformation and local translations in time respectively.  In
the orthogonal case one is lead to the following expression for
the currents
\begin{equation}
\vec{j} = -\left.{\partial H[\vec{A}_{p}] \over
\partial \vec{A}_{p}}\right|_{\vec{A}_{p} = 0}
\quad\mbox{and}\quad \vec{Q} = -\left.{\partial H[\vec{A}_{h}]
\over
\partial \vec{A}_{p}}\right|_{\vec{A}_{h} = 0},
\end{equation}
where $\vec{A}_{p}$ and $\vec{A}_{h}$ are gauge fields conjugate
to the currents and $H[\vec{A}_{p}]$ and $H[\vec{A}_{h}]$ denote
the gauged Hamiltonian, i.e.~the Hamiltonian with the
replacements $\vec{k} \rightarrow \vec{k}-\vec{A}_{p}$ and
$\vec{k} \rightarrow \vec{k}+\vec{A}_{h}\lrpart{\tau}$
respectively. This replacement is performed in both the kinetic
and the interaction terms but not in the field operators. In our
case however the overlap matrix appearing in the action depends
also on momentum and therefore the proper generalization of the
currents to non-orthogonal basis will also take the overlap matrix
into account.  Thus we obtain
\begin{eqnarray}
\vec{j} &=& -\frac{\partial
(O[\vec{A}_{p}]\lrpart{\tau}+H[\vec{A}_{p}])} {\partial
\vec{A}_{p}}|_{\vec{A}_{p} = 0} ,
\\
\vec{Q} &=& -\frac{\partial
(O[\vec{A}_{h}]\lrpart{\tau}+H[\vec{A}_{h}])} {\partial
\vec{A}_{h}}|_{\vec{A}_{h} = 0} .
\end{eqnarray}
Performing these operations leads to the following expressions
\begin{eqnarray}
\label{eq:particlecurrent} \vec{j} &=& \sum_{k\alpha\beta} \left(
\vec{v}_{k,\alpha\beta}B^{(0)}_{k,\alpha\beta}-
\vec{u}_{k,\alpha\beta}B^{(1)}_{k,\alpha\beta} \right) ,
\\
\label{eq:heatcurrent} \vec{Q} &=& \sum_{k\alpha\beta} \left(
\vec{v}_{k,\alpha\beta}B^{(1)}_{k,\alpha\beta}-
\vec{u}_{k,\alpha\beta}B^{(2)}_{k,\alpha\beta} \right) ,
\end{eqnarray}
where we have defined
\begin{equation}
  \label{eq:b_symbols}
  B_{k,\alpha\beta}^{(n)} =
  (-1)^{n}c_{k,\alpha}^{+}(\lrpart{\tau})^{n}c_{k,\beta} ,
\end{equation}
and
\begin{equation}
  \label{eq:velocities}
  \vec{v}_{k,\alpha\beta} =
  \frac{1}{\hbar}\grad{k}H^{0}_{k,\alpha\beta}
  \quad\mbox{and}\quad
  \vec{u}_{k,\alpha\beta} = \frac{1}{\hbar}\grad{k}O_{k,\alpha\beta},
\end{equation}
where $H^{0}_{k,\alpha\beta}$ is the tight-binding, LMTO
Hamiltonian of the system and $O_{k,\alpha\beta}$ is the overlap
matrix that captures the non-orthogonality of the basis that we
are using. The validity of the expressions above is not
restricted to DMFT and they are in fact true for all
density-density interactions such as the Hubbard interaction.
This is because the interaction terms are gauge invariant and
therefore they do not contribute to the expressions for the
currents.

\end{document}